\documentclass[preprint,12pt,authoryear]{elsarticle}

\usepackage{amssymb}
\usepackage{amsmath}
\usepackage{float}
\usepackage{lineno}
\usepackage{natbib}
\usepackage[hypertexnames=false]{hyperref}
\usepackage{siunitx}
\usepackage{xcolor}
\usepackage{longtable}
\usepackage{verbatim}
\usepackage{booktabs}
\usepackage{setspace}
\usepackage{array}
\journal{Earth and Planetary Science Letters}

\begin{document}

\begin{frontmatter}



\title{The role of salinity in Snowball Earth: A simplified coupled atmosphere--ocean--ice model} 


\author[label1]{Sounav Sengupta\corref{cor1}}
\ead{sounav.sengupta4@gmail.com}
\cortext[cor1]{Corresponding author.}
\author[label2]{Francisco E. Spaulding-Astudillo}
\author[label3]{Dorian S. Abbot}
\author[label4]{Eli Tziperman}
\author[label1]{Yosef Ashkenazy}


\affiliation[label1]{organization={Ben-Gurion University of the Negev},
            addressline={Sde Boker Campus},
            city={Midreshet Ben-Gurion},
            postcode={8499000},
            country={Israel}}

\affiliation[label2]{organization={University of California, Los Angeles},
            addressline={595 Charles E Young Dr E},
            city={Los Angeles},
            postcode={90095},
            state={California},
            country={USA}}

\affiliation[label3]{organization={The University of Chicago},
            addressline={5734 S. Ellis Avenue},
            city={Chicago},
            postcode={60637},
            state={Illinois},
            country={USA}}       

\affiliation[label4]{organization={Dept of Earth and Planetary Sciences and School of Engineering and Applied Sciences, Harvard University},
            addressline={20 Oxford Street},
            city={Cambridge},
            postcode={02138},
            state={Massachusetts},
            country={USA}}    



\begin{abstract}
The Neoproterozoic Snowball Earth events represent among the most extreme climate states in Earth's history, with ice covering all or nearly all of the ocean. Despite its control on the seawater freezing point and on the partitioning of water between the ocean and ice, salinity has not been well-represented in Snowball Earth models---typically prescribed, simplified, or ignored. Sea-glacier dynamics under open-ocean conditions have similarly lacked adequate treatment. We extend the one-dimensional coupled atmosphere–ocean–ice model of Pollard and Kasting (2005, J. Geophys. Res. Oceans 110, C07010) by adding prognostic latitudinal salinity dynamics, global ocean mass conservation and salt conservation, and a revised sea-glacier momentum equation, along with a greenhouse-gas term allowing transitions between full-snowball and partially open-ocean states. The model produces two equilibrium states exhibiting strong hysteresis under varying greenhouse forcing. Higher initial ocean salinity systematically lowers the CO$_2$ thresholds for both the onset and the termination of glaciation, reflecting freezing-point depression and coupled dynamics among salinity, ocean volume, and marine ice thickness. Under full glaciation, the shrinking ocean becomes hypersaline, feeding back on thermal structure and equilibrium ice thickness. Across a 72-case factorial ensemble spanning Neoproterozoic-relevant ranges, a marginal-effect analysis shows that only ice albedo shifts the thresholds more than initial salinity (69.1\% and 80.5\% versus 21.3\% and 9.7\% for initiation and deglaciation), followed by the solar constant (8.9\% and 5.6\%) and geothermal heat flux (0.6\% and 4.2\%). Salinity should therefore be treated as an active variable in studies of Snowball thresholds.

\end{abstract}



\begin{graphicalabstract}
    \centering
    \includegraphics[width=\textwidth]{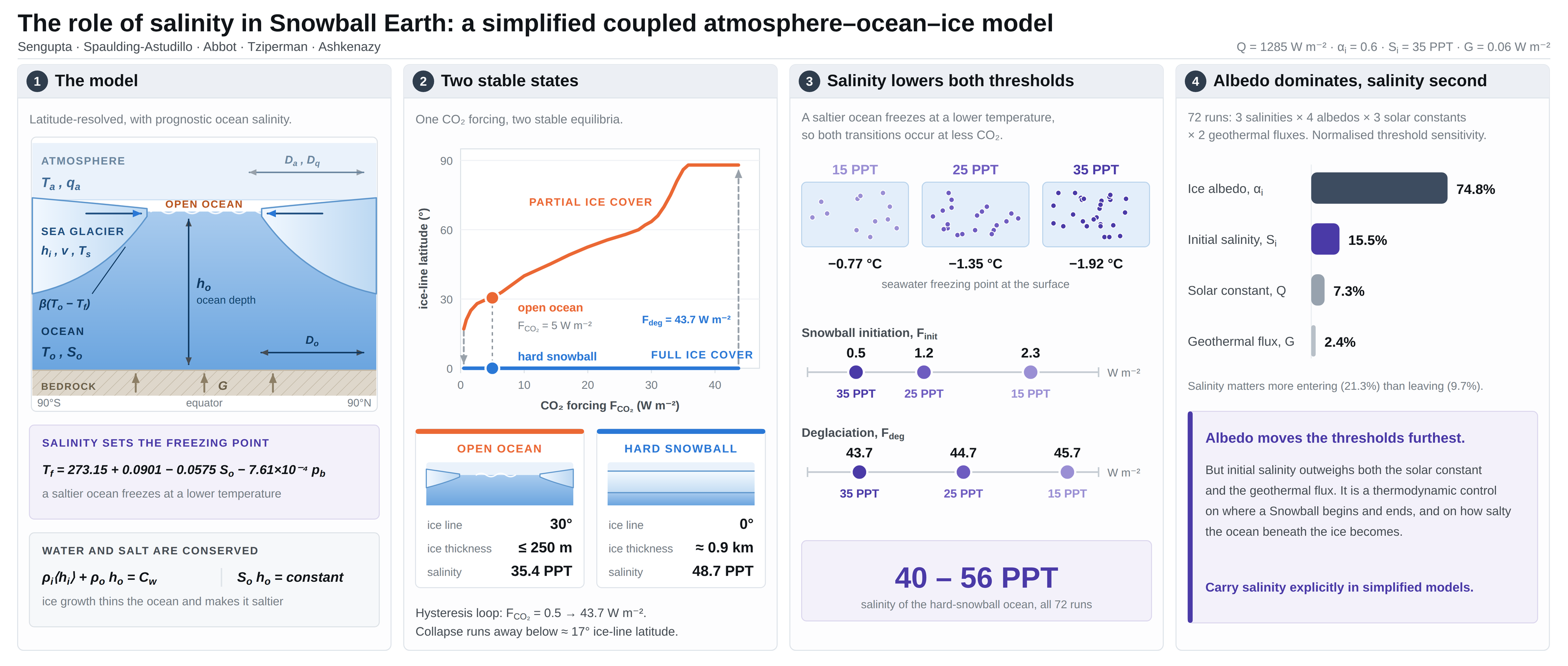}
\end{graphicalabstract}

\begin{highlights}
\item An idealized coupled Snowball Earth model includes prognostic ocean salinity.
\item Salt and water conservation couples ice growth, salinity, and ocean volume.
\item Higher salinity lowers the thresholds for glaciation and deglaciation.
\item Over Neoproterozoic ranges, salinity's effect is second only to ice albedo.
\end{highlights}


\end{frontmatter}



\section{Introduction}

During Snowball Earth episodes, ice extended to low latitudes and the oceans became globally or nearly globally ice-covered. Two major episodes are recognized in the Cryogenian Period, the Sturtian (ca.~717--659~Ma) and Marinoan (ca.~645--635~Ma) glaciations \citep{Rooney2015,Hoffman2017}. Evidence includes glacial diamictites and dropstones, paleomagnetic indications of glacial deposition at tropical paleolatitudes, and cap carbonates directly overlying glacial deposits \citep{Harland1964,Hoffman1998,HoffmanSchrag2002}, which motivated the Snowball Earth hypothesis of \citet{Kirschvink1992}. In its canonical form, low-latitude glaciation triggered a runaway ice--albedo feedback, and volcanic CO$_2$ accumulated over millions of years until greenhouse warming triggered deglaciation \citep{Hoffman1998,HoffmanSchrag2002}.



Whether the Cryogenian oceans were completely sealed beneath ice or retained an equatorial waterbelt remains debated \citep{Pierrehumbert2011,Hoffman2017}, and this question has motivated a wide hierarchy of models. Idealized and intermediate-complexity studies explored the roles of ocean heat transport, clouds, continental configuration, runoff, and CO$_2$ buildup in glaciation and deglaciation \citep{Hyde2000,Poulsen2004,Donnadieu2004,Pierrehumbert2005}, while general circulation models, sea-ice and shelf-ice models, and energy-balance frameworks examined full-snowball stability, Jormungand and waterbelt states, and deglaciation pathways \citep{Ashkenazy2013,Ashkenazy2014,Tziperman2012,Lewis2007,Abbot2011,Abbot2013,Rose2015,Horner2022}. Multiple equilibria between ice-covered and partially ice-covered states also arise in comprehensive climate models \citep{Marotzke2007}. Within this hierarchy, one-dimensional (meridional) models are particularly useful because they isolate the feedbacks governing climate hysteresis, following the energy-balance frameworks of \citet{Budyko1969} and \citet{Sellers1969}, later formalized by \citet{North1975,North1981}.

The present study builds on the model of \citet{Pollard2005}, who coupled a latitudinal energy-balance climate model to dynamically flowing marine ice and obtained equilibria with thin tropical ice and even persistent open water. This framework builds on \citet{Goodman2003}, who showed that floating marine-ice flow can homogenize ice thickness on planetary scales. \citet{Tziperman2012} derived the sea-glacier flow equations from the Stokes equations in spherical geometry, showing that even the latitude-only case acquires additional geometric and stress-divergence terms, and \citet{Pollard2017} asynchronously coupled zonal sea-glacier flow to a three-dimensional atmospheric general circulation model. Sea-glacier models have also been used to examine marine-ice invasion of restricted basins and potential refugia \citep{Tziperman2012,Campbell2014}. Unlike the comparatively thin sea ice driven by wind stress, ocean stress, and pack-ice rheology in most general circulation model studies \citep{Lewis2007,Horner2022}, a sea glacier is a thick floating viscous body governed by Glen-type flow, hydrostatic stress balance, and terminus force balance \citep{Pollard2005,Tziperman2012}, a more appropriate description once the ice reaches hundreds of meters to a kilometer in thickness.

Despite this progress, salinity remains comparatively underexplored in reduced Snowball Earth models, an omission that matters for two reasons. First, salinity lowers the freezing temperature of seawater, directly modifying the thermodynamic conditions for marine ice growth and melt. Second, sea-glacier growth reduces ocean volume; under a conserved salt inventory, this volume change necessarily alters the mean ocean salinity, which in turn feeds back on the freezing temperature.

Some earlier studies recognized this connection. \citet{Goodman2003} noted that meltwater near the ice margin can reduce surface salinity and raise the freezing temperature, and ocean-circulation studies under globally glaciated conditions showed that full-snowball oceans can have unusual circulation and density structures beneath thick ice \citep{Ashkenazy2013,Ashkenazy2014}. \citet{Olson2022} showed that ocean salinity strongly controls sea-ice extent and substantially shifts the Snowball glaciation threshold, and \citet{Liu2023} found that weakened ocean ventilation acts as a positive cooling feedback through additional CO$_2$ drawdown, facilitating Snowball onset. We do not address carbon-cycle or ventilation feedbacks; instead, we focus on how prognostic salinity, coupled with global ocean mass and salt conservation, modifies the hysteresis between partially open-ocean and full-snowball states and shifts both glaciation and deglaciation thresholds via freezing-point depression.

Salinity also affects seawater density and potential habitability \citep{Cullum2016}; both are discussed in Section~\ref{sec:discussion}.

Here we extend the one-dimensional atmosphere--ocean--sea-glacier model of \citet{Pollard2005} by incorporating prognostic latitudinal salinity, a salinity-dependent freezing temperature, and global ocean mass and salt conservation, with the single-layer ocean treated as an effective ocean column rather than a shallow mixed layer. The atmosphere--ocean component is coupled asynchronously to a floating sea-glacier model whose momentum balance retains a lateral-geometry correction associated with the Morland--MacAyeal shelf-flow formulation and the spherical stress-divergence correction of \citet{Tziperman2012}, which the simpler pressure-gradient balance of \citet{Pollard2005} omits. Ocean salinity therefore evolves with ice thickness and ocean volume rather than being prescribed.
 
Our aim is to quantify the effect of salinity on the glaciation and deglaciation thresholds; to provide context for the size of this effect, we also vary ice-surface albedo, solar constant, and geothermal heat flux. Ice albedo sets the strength of the ice--albedo feedback, a major uncertainty in Snowball stability and deglaciation \citep{Lewis2006,Abbot2010,Abbot2011}; the solar constants used are 5.0--6.0\% below the present-day value of 1367~\(\mathrm{W\,m^{-2}}\), consistent with the fainter Neoproterozoic Sun \citep{Gough1981}; and geothermal heat flux sets the basal heat supply beneath thick marine ice \citep{JaupartMareschal2007,Mareschal2013}. We quantify the relative influence of these controls with a marginal-effect analysis of the glaciation and deglaciation thresholds.

Section~\ref{sec:model} describes the model, Section~\ref{sec:results} presents the bistable equilibria and their sensitivity to salinity, albedo, and the other parameters, Section~\ref{sec:discussion} discusses the implications, and Section~\ref{sec:summary} summarizes the conclusions.



\section{Model description}
\label{sec:model}


\subsection{Compact atmospheric and oceanic equations}

The model follows the Pollard--Kasting hierarchy of a diffusive atmosphere--slab-ocean system coupled to floating marine ice \citep{Pollard2005}, with a prognostic salinity equation and a salinity-dependent freezing temperature. The planet is an aquaplanet without continents or rotation, forced by annual-mean insolation (configuration and numerical settings in \ref{app:configuration}). Parameter values follow \citet{Pollard2005}, except for a darker ocean albedo (\(\alpha_o=0.10\) rather than 0.15) and an ice--ocean heat-exchange coefficient \(\beta=400~\mathrm{W\,m^{-2}\,K^{-1}}\) \citep{Ashkenazy2018,HollandJenkins1999}. We write the thermodynamic equations in compact form on a one-dimensional co-latitude (\(\theta\)) domain; the full spherical-coordinate equations, flux parameterizations, numerical scheme, and symbol definitions are given in Appendix~A (Tables~\ref{tab:variables} and~\ref{tab:parameters}). The prognostic variables are ocean temperature \(T_o\) and salinity \(S_o\), atmospheric temperature \(T_a\) and specific humidity \(q_a\), and ice-surface temperature \(T_s\); \(h_o\) is the globally averaged effective ocean depth, \(h_a\) and \(h_q\) are effective atmospheric layer thicknesses, \(C_s\) is the heat capacity of the snow--ice surface layer, and \(p_b\) is the basal pressure.

The meridional diffusion terms use the spherical operator
\begin{equation}
\mathcal{D}_{j}[X_j]
=
\frac{1}{r^{2}\sin\theta}
\frac{\partial}{\partial\theta}
\left(
D_j \sin\theta \frac{\partial X_j}{\partial\theta}
\right),
\label{eq:diffusion_operator}
\end{equation}
where \((X_j,D_j)\) is \((T_o,D_o)\), \((T_a,D_a)\), \((q_a,D_q)\), or \((S_o,D_o)\) for \(j=o,a,q,S\); ocean temperature and salinity share the diffusivity \(D_o\).

The atmospheric temperature equation is
\begin{equation}
\rho_a c_a h_a \frac{\partial T_a}{\partial t}
=
\rho_a c_a h_a \,\mathcal{D}_{a}[T_a]
+
\mathcal{F}_{a}(T_a,T_s,T_o,q_a)
+
F_{\mathrm{CO_2}},
\label{eq:Ta_compact_main}
\end{equation}
where \(\mathcal{F}_{a}\) is the net radiative, sensible, and latent atmospheric forcing (Eq.~\eqref{eq:app_Ta}) and \(F_{\mathrm{CO_2}}\) is a spatially uniform longwave heating term that is varied to trace the hysteresis curves.

The atmospheric specific-humidity, $q_a$, equation is
\begin{equation}
\rho_a h_q \frac{\partial q_a}{\partial t}
=
\rho_a h_q \,\mathcal{D}_{q}[q_a]
-
\mathcal{M}(T_a,T_s,T_o,q_a),
\label{eq:qa_compact_main}
\end{equation}
where \(\mathcal{M}\equiv P-E\) (Eq.~\eqref{eq:app_qa}) is precipitation minus evaporation or sublimation.

The ocean-temperature equation is
\begin{equation}
\rho_o c_o h_o \frac{\partial T_o}{\partial t}
=
\rho_o c_o h_o \,\mathcal{D}_{o}[T_o]
+
\mathcal{F}_{o}(T_o,T_a,T_s,q_a,S_o,h_o,h_i)
-
\beta\left(T_o-T_f\right),
\label{eq:To_compact_main}
\end{equation}
where \(\mathcal{F}_{o}\) is the net radiative, turbulent, and geothermal oceanic forcing (Eq.~\eqref{eq:app_To}), and \(-\beta(T_o-T_f)\) is a linear ice--ocean heat exchange that removes heat from the ocean when \(T_o>T_f\), promoting basal melting, and changes sign, corresponding to basal freezing, when the ocean is colder than the local freezing temperature.

The salinity equation is written as
\begin{equation}
h_o \frac{\partial S_o}{\partial t}
=
h_o \mathcal{D}_{S}[S_o]
+
\mathcal{B}_{S}(S_o,F_o,E,P),
\label{eq:So_compact_main}
\end{equation}
where \(\mathcal{B}_{S}\) is the local salinity source term defined as
\begin{equation}
\mathcal{B}_{S}(S_o,F_o,E,P)
=
\langle S_o \rangle F_o
+
\frac{\langle S_o \rangle}{\rho_o} (E-P),
\label{eq:BS_compact}
\end{equation}where \(\langle S_o \rangle\) is the mean ocean salinity, following the salinity-transport formulation of \citet{Ashkenazy2018}; \(F_o\) and \((E-P)/\rho_o\) are freshwater-equivalent depth tendencies (\(\mathrm{m\,s^{-1}}\)). Because latitudinal salinity variations are small compared with the difference in mean salinity between the open-ocean and full-snowball equilibria (Section~\ref{sec:bistability}), using \(\langle S_o\rangle\) in the source term does not materially affect the salt balance; the prognostic equation still sets the local salinity and hence the local freezing temperature, ocean temperature, and ice thickness. Global salt closure is additionally enforced after each coupling step.

Here \(P\) and \(E\) are the Pollard--Kasting water-mass fluxes (\(\mathrm{kg\,m^{-2}\,s^{-1}}\)), and the basal freshwater source or sink associated with freezing or melting at the ice--ocean interface is
\begin{equation}
F_o
=
\frac{1}{L \rho_o}
\left[
\kappa_i \frac{T_f-T_s}{h_i}
-
\beta (T_o-T_f)
\right],
\label{eq:Fo_compact}
\end{equation}
where \(L\) is the latent heat of fusion. The salinity budget is thus coupled to basal freezing and melting and to the net freshwater flux at the open-ocean surface. 

Following \citet{Pollard2005}, the surface temperature of snow and ice, $T_s$, is written as
\begin{equation}
C_s \frac{\partial T_s}{\partial t}
=
\mathcal{F}_{s}(T_s,T_a,q_a)
+
F_c(T_s,T_f,h_i),
\label{eq:Ts_compact_main}
\end{equation}
where \(\mathcal{F}_{s}\) (Eq.~\eqref{eq:app_Ts}) is the net radiative and turbulent forcing at the snow--ice surface, and the conductive flux through ice of thickness \(h_i\) is
\begin{equation}
F_c = \kappa_i \frac{T_f-T_s}{h_i},
\label{eq:Fc_compact}
\end{equation}
where \(\kappa_i\) is the thermal conductivity of ice.

The freezing point of seawater decreases with salinity and pressure; we use the standard linear liquidus approximation based on \citet{Millero1978} and commonly used in ice-shelf/ocean boundary-layer models \citep{HollandJenkins1999,HellmerOlbers1989,JenkinsHellmerHolland2001,Losch2008}. Expressed in Kelvin,
\begin{equation}
T_f =
273.15 + 0.0901 - 0.0575\,S_o - 7.61\times10^{-4}p_b,
\label{eq:Tf_linear_main}
\end{equation}
where \(S_o\) is in PPT and \(p_b\) is the basal pressure in dbar. Through \(T_f\), salinity variations alter the ice--ocean heat exchange (Eq.~\eqref{eq:To_compact_main}), basal freezing and melting (Eq.~\eqref{eq:Fo_compact}), and the conductive flux through the ice (Eq.~\eqref{eq:Ts_compact_main}).

\subsection{Mass balance and global salt closure}
\label{sec:mass_salt_closure}

In most slab-ocean models, \(h_o\) is the depth of a 50--100~m mixed layer \citep{Ferreira2011}. Here, \(h_o\) is instead the effective depth of the whole ocean column, because a globally glaciated ocean is expected to be weakly stratified and relatively well mixed, consistent with studies that emphasize vigorous transport and efficient mixing beneath thick ice \citep{Ashkenazy2013,Ashkenazy2014,Ashkenazy2018,Yang2012}.

Changes in ocean volume must then be coupled consistently to changes in ice volume. Neglecting the small atmospheric water reservoir, the model conserves the total water inventory through the global-mean mass balance
\begin{equation}
    \rho_i \langle h_i \rangle + \rho_o h_o = C_w,
    \label{eq:massclosure_epsl}
\end{equation}
where \(\rho_i\) and \(\rho_o\) are the densities of ice and seawater, respectively, \(\langle h_i \rangle\) is the area-weighted global-mean ice thickness, and \(h_o\) is the globally averaged effective ocean thickness (equivalently, depth). The angular brackets denote
\begin{equation}
\langle h_i \rangle = \frac{1}{2}\int_{0}^{\pi} h_i(\theta,t)\sin\theta\,d\theta,
\label{eq:mean_hi_epsl}
\end{equation}
and \(C_w\) is the conserved column water mass per unit area; only the global mean of the local ice thickness \(h_i(\theta,t)\) enters this bookkeeping. 

As in the Pollard--Kasting framework, \(\rho_i\) and \(\rho_o\) are constant reference densities. Because buoyancy-driven circulation is not resolved in a slab model, salinity acts only through the freezing-temperature relation and the salt-conservation closure.

Following Eq.~\eqref{eq:massclosure_epsl}, variations in the uniform ocean depth are related to changes in the global-mean ice thickness through
\begin{equation}
\delta h_o = -\frac{\rho_i}{\rho_o}\,\delta\langle h_i \rangle.
\label{eq:dho_epsl}
\end{equation}

Because the atmosphere--ocean equations are integrated with the ice thickness held fixed, and the ice-flow equations with the atmospheric and oceanic fields held fixed (Section~\ref{sec:coupling}), the salinity equation alone does not guarantee salt conservation across the coupled integration. We therefore impose a global salt closure after each coupling step,
\begin{equation}
\langle S_o \rangle_i h_{o,i} = \langle S_o \rangle_f h_{o,f},
\label{eq:saltclosure_epsl}
\end{equation}
where the subscripts \(i\) and \(f\) denote values before and after the correction. In fully glaciated states, marine-ice thickness is controlled primarily by geothermal heat flux \citep{Hyde2000,Pierrehumbert2005,Pollard2005}; as the ice thickens, the ocean depth decreases and the conserved salt inventory is concentrated into a smaller volume, lowering the freezing temperature (Eq.~\eqref{eq:Tf_linear_main}) and feeding back on the ocean thermal state and the conditions required for deglaciation.

\subsection{Modified 1D sea-glacier momentum equation}

The ice flow follows the one-dimensional spherical sea-glacier formulation of \citet{Tziperman2012}, whose latitude-only meridional momentum balance is
\begin{equation}
\begin{split}
&\frac{1}{\sin\theta}
\frac{\partial}{\partial\theta}
\left(
B \sin\theta \frac{\partial v}{\partial\theta}
\right)
+
\frac{\partial}{\partial\theta}
\left[
\frac{B}{\sin\theta}
\frac{\partial}{\partial\theta}
\left(
v\sin\theta
\right)
\right]
\\
&\quad -
\cot^{2}\theta\,Bv
-
g\rho_i(1-\mu)\,h_i\frac{\partial h_i}{\partial\theta}
= 0,
\end{split}
\label{eq:momentum_epsl}
\end{equation}
where \(\theta\) is the co-latitude, \(v\) is the meridional ice velocity, \(B\) is the effective ice viscosity, and \(\mu=\rho_i/\rho_o\). Relative to the Pollard--Kasting formulation, Eq.~\eqref{eq:momentum_epsl} retains the spherical stress-divergence correction derived by \citet{Tziperman2012}, represented by the \(-\cot^{2}\theta\,Bv\) term. The first two terms are the spherical membrane-stress divergence, and the final term is the hydrostatic driving stress from gradients in ice thickness. The associated terminus force balance is given in~\ref{app:sea_glacier_equation}, Eq.~\eqref{eq:app_front}.
The corresponding local ice-thickness evolution equation is
\begin{equation}
    \frac{\partial h_i}{\partial t}+\frac{1}{r \sin{\theta}}\frac{\partial}{\partial \theta}(v h_i\sin{\theta})=\kappa\nabla^2h_i+S(\theta,t),
    \label{eq:ice-thickness}
\end{equation}
where the left-hand side represents local thickening plus the divergence of meridional ice flux. The diffusion term is included only for numerical regularization, so the coefficient \(\kappa\) is chosen as small as stable integration permits (for \(\kappa=1~\mathrm{m^2\,s^{-1}}\), a representative equilibrium calculation with \(Q=1285~\mathrm{W\,m^{-2}}\) and \(F_{\mathrm{CO_2}}=-1~\mathrm{W\,m^{-2}}\) yields an area-weighted mean absolute tendency of approximately \(4\%\) of the combined magnitudes of the ice-flux-divergence and thermodynamic source terms). The source term \(S(\theta,t)\), with units of \(\mathrm{m\,s^{-1}}\), is the net local thermodynamic ice-thickness tendency from basal freezing or melting and surface accumulation or ablation.

Calving-front migration is represented by a subgrid fractional ice-cover parameterization based on \citet{Albrecht2011} (\ref{app:subgrid_front}).

Prognostic equations are integrated with a second-order Adams--Bashforth scheme, whereas the momentum equation is solved diagnostically and iteratively at each coupling step, because the effective viscosity depends on strain rate through Glen's flow law and on temperature through \(A(T)\) (\ref{app:numerical_implementation}).

\subsection{Asynchronously coupled framework}
\label{sec:coupling}

In each coupling cycle, the atmosphere--ocean module is integrated to convergence under a prescribed ice geometry, and the resulting fields force the sea-glacier model, which updates the ice thickness and front position on the slower ice-flow timescale. After each ice update, the ocean depth and mean salinity are updated through the water-mass and salt closures (Section~\ref{sec:mass_salt_closure}), and the updated geometry, salinity, and freezing temperature are passed back to the atmosphere--ocean module. The cycle is repeated until the model reaches a steady state.


\section{Results}
\label{sec:results}


\subsection{Bistability and contrasting equilibrium states}
\label{sec:bistability}


We next examine the model equilibria obtained by varying the imposed CO$_2$ radiative forcing, \(F_{\mathrm{CO_2}}\). To identify possible multiple stable states, each forcing value is initialized from two contrasting initial conditions: an almost ice-free state and a fully ice-covered state. Both integrations are continued until they reach steady state. Near both the Snowball initiation and deglaciation transitions, $F_{\mathrm{CO_2}}$ was varied in increments of $0.1~\mathrm{W\,m^{-2}}$ to resolve the transition thresholds. At sufficiently low forcing, both initial conditions converge to a full-snowball state, whereas at sufficiently high forcing they converge to an almost ice-free open-ocean state. Between these limits, the two initial conditions can converge to different stable equilibria, defining the hysteresis loop.

\begin{figure}[htbp]
    \centering
    \includegraphics[width=0.82\textwidth]{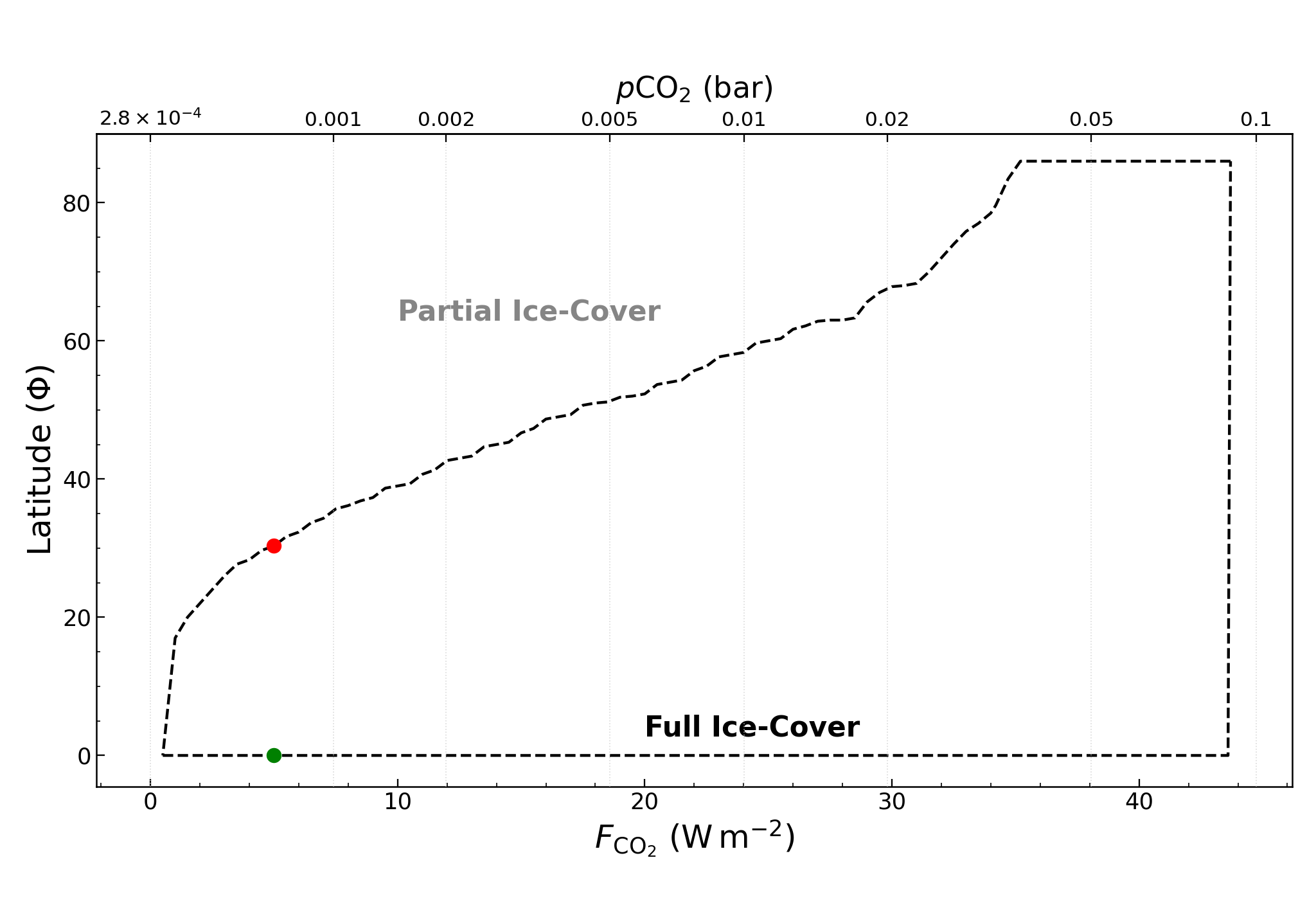}
    \caption{
    Ice-line hysteresis for the reference experiment with solar constant 
    $Q=1285~\mathrm{W\,m^{-2}}$, initial salinity 
    $S_i=35~\mathrm{PPT}$, ice-surface albedo $\alpha_i=0.6$, and geothermal heat flux 
    $G=0.06~\mathrm{W\,m^{-2}}$. The curve shows the equilibrium ice-line latitude as a function of imposed CO$_2$ radiative forcing, $F_{\mathrm{CO_2}}$. The red and green opaque circles indicate two distinct stable equilibria at the same forcing, $F_{\mathrm{CO_2}}=5~\mathrm{W\,m^{-2}}$, whose associated fields are shown in Fig.~\ref{fig:bistable_profiles}.
    }
    \label{fig:hysteresis_comp}
\end{figure}


The reference experiment, shown in Fig.~\ref{fig:hysteresis_comp}, exhibits a clear hysteresis loop: within a finite range of imposed CO$_2$ forcing, the equilibrium climate depends on the initial condition. As a representative point within this bistable regime, we focus on 
$F_{\mathrm{CO_2}}=5~\mathrm{W\,m^{-2}}$, for which the model admits two stable equilibria: an open-ocean state and a full-snowball state.

\begin{figure}[htbp]
    \centering
    \includegraphics[width=0.95\textwidth]{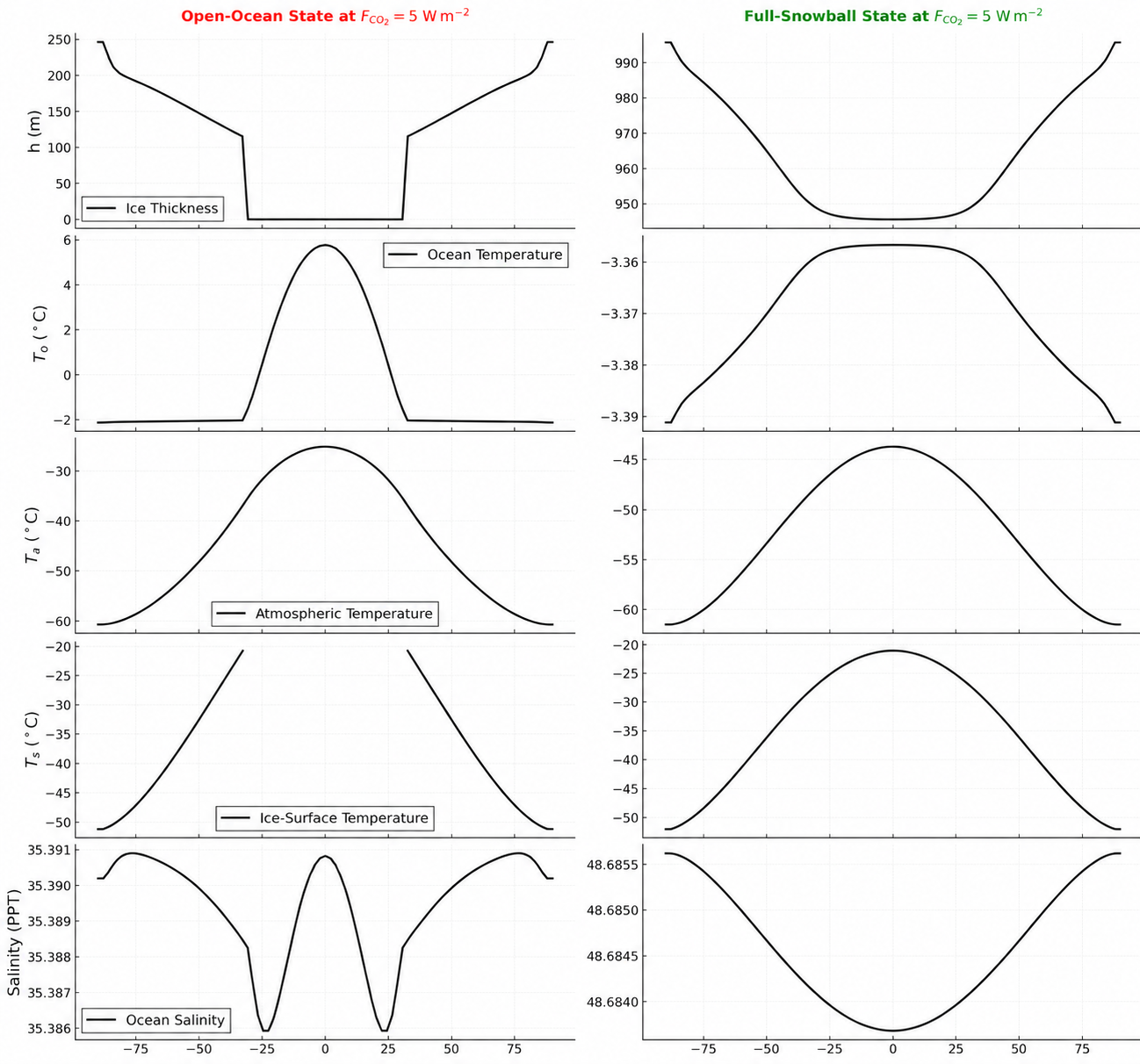}
    \caption{
    Latitudinal dependence of the two bistable climate states marked in Fig.~\ref{fig:hysteresis_comp} at 
    $F_{\mathrm{CO_2}}=5~\mathrm{W\,m^{-2}}$. The left column shows the open-ocean (partial ice cover) equilibrium, while the right column shows the full-snowball equilibrium. The panels compare ice thickness, ocean temperature, atmospheric temperature, ice-surface temperature, and ocean salinity.
    }
    \label{fig:bistable_profiles}
\end{figure}

The latitudinal structure of the full-snowball ice thickness is broadly consistent with the reference solution of \citet{Pollard2005}, with values of order \(1~\mathrm{km}\) in both studies. The no-flow conductive estimate below provides a reference thermodynamic thickness scale for the full-snowball branch and gives an order-of-magnitude thickness near \(1~\mathrm{km}\) for the reference parameters.
\begin{equation}
\overline{h_i}
\approx
\frac{\kappa_i\,\overline{(T_f-T_s)}}{G},
\label{eq:h-noflow}
\end{equation}
where \(\overline{h_i}\) is the global-mean ice thickness, \(\kappa_i\) is the thermal conductivity of ice, \(\overline{(T_f-T_s)}\) is the global-mean basal-to-surface temperature difference through the ice, and \(G\) is the geothermal heat flux. This relation provides a leading-order, no-flow thickness scale for the globally ice-covered branch \citep{Hyde2000,Pierrehumbert2005,Pollard2005}. In a numerical calculation with the ice-flow term disabled and negligible ocean diffusivity, its local form, using \(T_f(\theta)\) and \(T_s(\theta)\), gives the conductive ice-thickness profile. In the full-snowball profile in Fig.~\ref{fig:bistable_profiles}, lateral ice redistribution homogenizes the ice cover, so Eq.~\eqref{eq:h-noflow} should be interpreted principally as an estimate of the global mean. The local thickness also has smaller departures from the surface mass balance, including precipitation and evaporation or sublimation. It is not intended to describe the thinner, dynamically flowing, and spatially restricted ice of the open-ocean branch, where the tropical ocean remains ice-free.

The salinity fields (bottom panels of Fig.~\ref{fig:bistable_profiles}) also differ strongly between the two equilibrium states. In the full-snowball state, the larger equilibrium ice thickness reduces the ocean volume and increases the mean salinity to \(48.68~\mathrm{PPT}\), compared with \(35.38~\mathrm{PPT}\) in the open-ocean state. The resulting freezing-point depression is reflected in the ocean-temperature panels: the full-snowball ocean is about \(-3.37^\circ\mathrm{C}\), whereas the partially covered ocean is warmer than approximately \(-2^\circ\mathrm{C}\) over much of the open-water region. Beyond this mean shift, the latitudinal salinity structure also changes qualitatively. In the full-snowball state, salinity variations are small and broadly follow the ice-thickness distribution, reflecting the coupled evolution of ice thickness, ocean volume, and freezing temperature. In the open-ocean state, by contrast, salinity peaks at the equator where evaporation is strongest over exposed ocean, reaches a minimum near the ice edge where melting is strongest, and develops broader high-latitude peaks where freezing rejects salt. Thus, the two equilibria develop markedly different salinity profiles despite having the same external forcing. Larger ocean diffusivity or weaker effective ice stiffness, which permits stronger ice redistribution, would tend to smooth these gradients; lower diffusivity or stronger effective ice stiffness would allow sharper contrasts.


The corresponding oceanic and atmospheric temperature profiles reinforce this interpretation. The open-ocean state maintains a relatively warm equatorial ocean and atmosphere, whereas the full-snowball state remains substantially colder across the domain. In the full-snowball state, the high ice albedo reduces absorbed shortwave radiation and maintains low atmospheric temperatures \citep{Schneider1992,Pollard2005}. The ocean temperature varies only weakly with latitude because the thick, continuous ice cover insulates the ocean from direct surface forcing and ocean diffusion smooths the residual temperature gradients. By contrast, atmospheric and ice-surface temperatures retain stronger equator-to-pole gradients because they respond directly to the latitudinal pattern of absorbed solar radiation and exchange longwave and turbulent heat with each other. Within the open-ocean region, ocean temperatures remain above the local freezing temperature because of direct shortwave absorption and the ocean's large heat capacity, while meridional ocean transport carries heat away from low latitudes.

\subsection{Sensitivity to initial salinity}


\begin{figure}[htbp]
    \centering
    \includegraphics[width=0.95\textwidth]{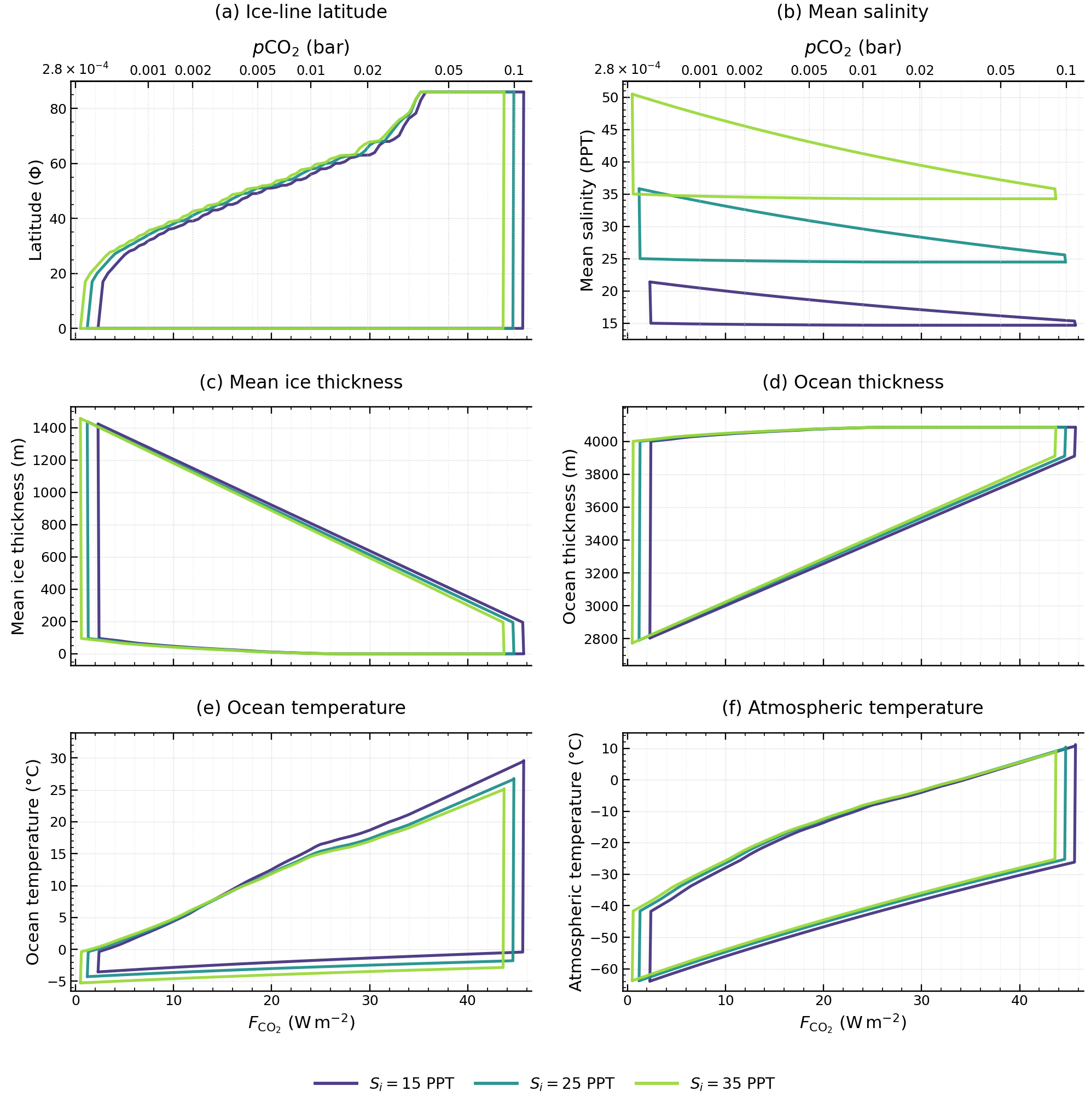}
    \caption{
    Hysteresis curves of global-mean model variables for different initial salinities. The model parameters are fixed at $\alpha_i=0.6$, $Q=1285~\mathrm{W\,m^{-2}}$, and $G=0.06~\mathrm{W\,m^{-2}}$, while the initial salinity is varied between $S_i=15$, $25$, and $35~\mathrm{PPT}$. Panels show (a) ice-line latitude, (b) mean salinity, (c) mean ice thickness, (d) mean ocean depth, (e) ocean temperature, and (f) atmospheric temperature. The lower x-axis indicates the greenhouse forcing $F_{\mathrm{CO_2}}$ in $\mathrm{W\,m^{-2}}$, while the upper x-axis in panels (a) and (b) shows the CO$_2$ concentration in bar on a logarithmic scale. Thus, the deglaciation transitions on the right are substantially separated when expressed in terms of CO$_2$ concentration. The salinity, ice-thickness, and mean-ocean-depth panels highlight the implemented salt and mass conservation: changes in ice volume alter the ocean volume, which in turn modifies the mean ocean salinity during transitions between open-ocean and snowball states.
    }
    \label{fig:salinity_comparison_subplots}
\end{figure}


The effect of initial salinity on the hysteresis curve is shown in Fig.~\ref{fig:salinity_comparison_subplots}a. Variations in the initial salinity systematically shift the values of $F_{\mathrm{CO_2}}$ associated with both Snowball initiation and deglaciation. Specifically, fresher oceans require larger greenhouse-gas forcing to deglaciate and transition into an open-ocean state, while the transition to global glaciation also occurs at higher forcing levels. This behavior follows directly from the salinity dependence of the seawater freezing temperature: increasing salinity lowers the freezing point, making the sea glacier more difficult to form and easier to melt. Salinity, therefore, acts as an active control on the greenhouse-forcing thresholds governing transitions between the snowball states.

This sensitivity to initial salinity is further illustrated in Fig.~\ref{fig:salinity_comparison_subplots}, which shows the hysteresis curves of several global-mean variables for the same set of initial salinities. For the full-snowball branch, increasing greenhouse-gas forcing enhances sea-glacier melting, increases the ocean volume, and reduces the mean ocean salinity (upper branch of Fig.~\ref{fig:salinity_comparison_subplots}b). This qualitative behavior is consistent across all initial salinity values. In contrast, under open-ocean conditions the mean salinity remains comparatively stable and close to its initial value because the ocean reservoir is much larger and the marine-ice volume is much smaller. This behavior explains the salinity-dependent sensitivity of the deglaciation threshold: once the system enters a full-snowball state, the reduced ocean volume produces a higher equilibrium salinity, which further depresses the seawater freezing temperature and changes the greenhouse forcing required for retreat of the global ice cover.



As shown in Fig.~\ref{fig:salinity_comparison_subplots}d, as marine-ice volume increases, the effective ocean depth decreases, concentrating the dissolved salt into a progressively smaller volume and thereby increasing the mean salinity. Consequently, the ocean-depth and salinity curves evolve in opposite directions and together illustrate the internal consistency of the coupled system: salinity changes arise naturally from the dynamically linked evolution of ice thickness, ocean volume, and global salt conservation.

The quantitative summary of these salinity-sensitivity experiments is given in Table~\ref{tab:S_salinity_sensitivity}. Increasing the initial salinity from \(15\) to \(35~\mathrm{PPT}\) lowers the Snowball-initiation threshold from \(2.3\) to \(0.5~\mathrm{W\,m^{-2}}\) and lowers the deglaciation threshold from \(45.7\) to \(43.7~\mathrm{W\,m^{-2}}\). In the approximate CO$_2$ scale shown on the upper axis of Fig.~\ref{fig:salinity_comparison_subplots}a, based on the high-CO$_2$ radiative-forcing parameterization of \citet{ByrneGoldblatt2014}, the corresponding deglaciation estimate decreases from \(0.110\) to \(0.09~\mathrm{bar}\). Thus, the salinity dependence of the deglaciation threshold is physically meaningful even though the forcing shift is only about \(2~\mathrm{W\,m^{-2}}\) across the sampled salinity range. Table~\ref{tab:S_salinity_sensitivity} also shows that the final salinity increases substantially in the full-snowball state as the ocean volume is reduced, while the ratio \(S_i/S_f\) remains nearly constant across the three cases. This indicates that salinity amplification is controlled mainly by the common mass-conservation constraint imposed by ice growth, rather than by the absolute initial salinity alone. The threshold shifts can be related directly to freezing-point depression (Eq.~\eqref{eq:Tf_linear_main}). On the open-ocean branch, the mean salinity remains close to its initial value, so increasing \(S_i\) from 15 to \(35~\mathrm{PPT}\) lowers \(T_f\) by about \(1.15~\mathrm{K}\) and the initiation threshold by \(1.8~\mathrm{W\,m^{-2}}\), approximately \(1.6~\mathrm{W\,m^{-2}}\) per kelvin of freezing-point depression. Just before deglaciation, the full-snowball salinities of 21.4--50.5~PPT differ in \(T_f\) by about \(1.7~\mathrm{K}\), and the deglaciation threshold shifts by \(2.0~\mathrm{W\,m^{-2}}\), approximately \(1.2~\mathrm{W\,m^{-2}}\) per kelvin.

\begin{table}[htbp]
\centering
\caption{Summary of glaciation and deglaciation transitions for different initial salinities; the other parameters are \(\alpha_i=0.6\), \(Q=1285\,\mathrm{W\,m^{-2}}\), and \(G=0.06\,\mathrm{W\,m^{-2}}\). Here, \(S_f\) denotes the global-mean ocean salinity immediately before the deglaciation transition in the full-snowball equilibrium. Approximate CO$_2$ concentrations are obtained using the \citet{ByrneGoldblatt2014} relation, \(F_{\mathrm{CO_2}}=5.32\ln(C/C_0)+0.39[\ln(C/C_0)]^2\), with \(C_0=2.8\times10^{-4}\,\mathrm{bar}\). Deglaciation concentrations above \(0.01\,\mathrm{bar}\) are extrapolations beyond the calibrated range of the simplified relation and are therefore reported as qualitative estimates.}
\label{tab:S_salinity_sensitivity}
\footnotesize
\setlength{\tabcolsep}{4pt}
\renewcommand{\arraystretch}{1.12}
\begin{tabular}{cccccccc}
\toprule
\(S_i\) & \(\overline{h_i}\) & \(S_f\) & \(F_{\mathrm{init}}\)
& \(p\mathrm{CO}_{2,\mathrm{init}}\) & \(F_{\mathrm{deg}}\)
& \(p\mathrm{CO}_{2,\mathrm{deg}}\) & \(S_i/S_f\) \\
PPT & \(\mathrm{m}\) & \(\mathrm{PPT}\) & \(\mathrm{W\,m^{-2}}\)
& \(\mathrm{bar}\) & \(\mathrm{W\,m^{-2}}\) & \(\mathrm{bar}\) & -- \\
\midrule
15 & 1423.4 & 21.39 & 2.3 & \(4.26\times10^{-4}\) & 45.7 & 0.110 & 0.70 \\
25 & 1437.9 & 35.82 & 1.2 & \(3.50\times10^{-4}\) & 44.7 & 0.0995 & 0.70 \\
35 & 1457.8 & 50.48 & 0.5 & \(3.07\times10^{-4}\) & 43.7 & 0.0899 & 0.69 \\
\bottomrule
\end{tabular}
\end{table}

The ocean-temperature and atmospheric-temperature hysteresis curves also exhibit distinct warm and cold branches that co-vary with the salinity and ocean-depth branches, and their overall shape remains qualitatively similar across initial salinities. On the full-snowball branch, the mean ice thickness, ocean depth, and ocean and atmospheric temperatures vary approximately linearly with the applied greenhouse-heating term \(F_{\mathrm{CO_2}}\), corresponding to increasing atmospheric \(p\mathrm{CO}_2\) concentration; mean salinity shows a weaker but similar tendency. The separation between the warm and cold branches is strongly variable dependent: it changes markedly for ocean temperature but more moderately for atmospheric temperature. These simple branch relationships provide a useful diagnostic for anticipating the evolution of the hysteresis curves once the glaciation and deglaciation thresholds are identified. Taken together, the panels in Fig.~\ref{fig:salinity_comparison_subplots} show that salinity influences both the entry into and escape from Snowball conditions through its control on the freezing temperature and its coupling to the evolving ocean volume.

\subsection{Sensitivity to ice-surface albedo}

\begin{figure}[htbp]
    \centering
    \includegraphics[width=0.95\textwidth]{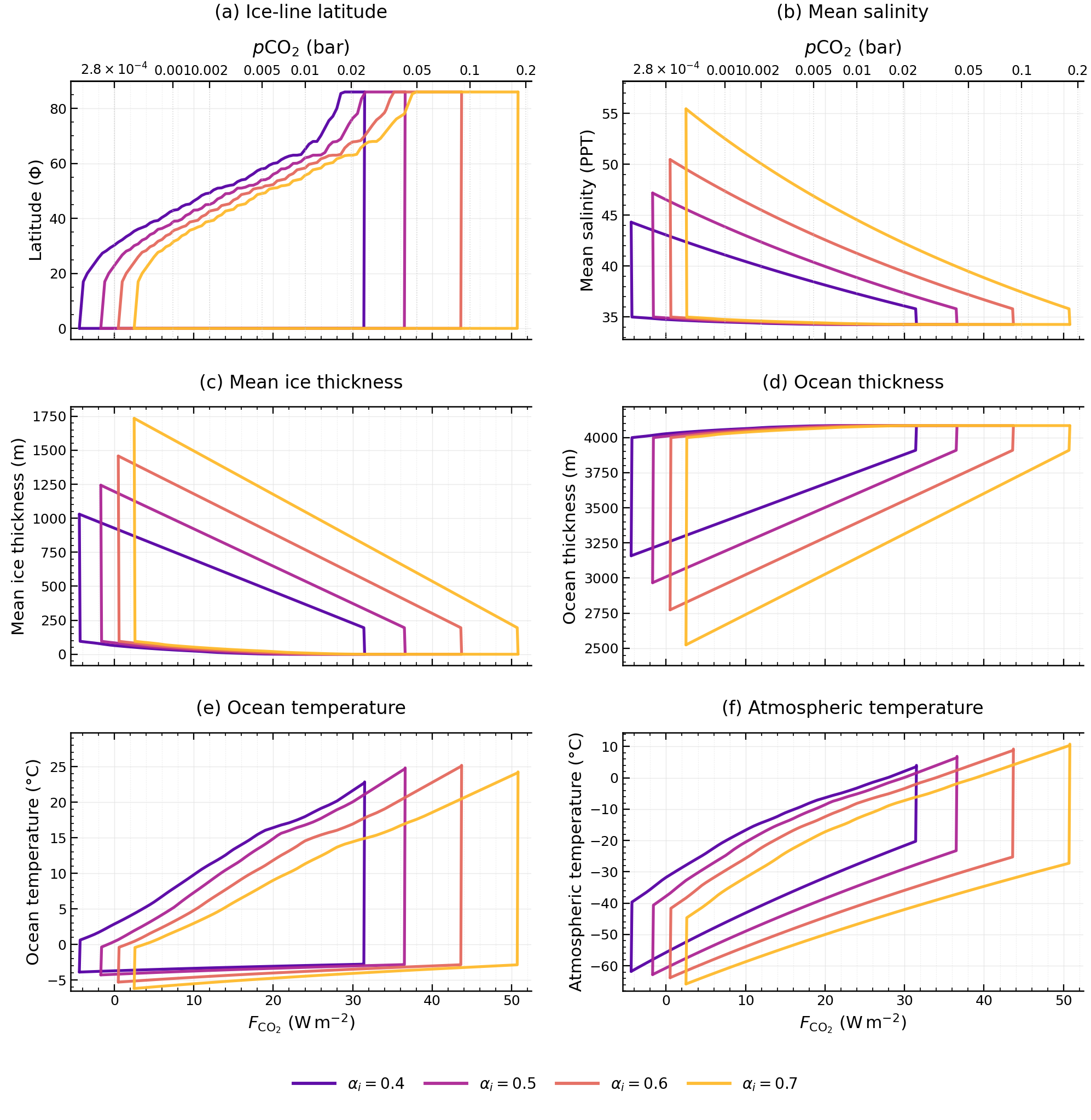}
    \caption{
    Dependence of hysteresis curves on ice albedo. The experiments use an initial salinity
    $S_i=35~\mathrm{PPT}$, solar constant $Q=1285~\mathrm{W\,m^{-2}}$, and geothermal heat flux
    $G=0.06~\mathrm{W\,m^{-2}}$, while the ice albedo is varied over
    $\alpha_i=0.4$, $0.5$, $0.6$, and $0.7$. (a) Ice-line latitude, (b) mean salinity, (c) mean ice thickness, (d) mean ocean depth, (e) ocean temperature, and (f) atmospheric temperature are shown as functions of greenhouse forcing $F_{\mathrm{CO_2}}$ (lower x-axis) and CO$_2$ concentration (upper x-axis). Higher ice albedo widens the hysteresis curves and shifts them toward larger greenhouse forcing, while the coupled salinity, ice-thickness, and mean-ocean-depth changes reflect the conservation of salt and water mass during ice expansion and retreat.
    }
    \label{fig:albedo_intercomparison}
\end{figure}

Although the ice--albedo feedback is already known to be important, we examine it here to compare its influence with that of salinity, the main focus of this study. Variations in ice albedo produced the largest impact on both the glaciation and deglaciation thresholds; see Figs.~\ref{fig:albedo_intercomparison} and~\ref{fig:sensitivity_weightage}. This result is expected, because ice albedo directly controls the strength of the ice--albedo feedback, which has long been recognized as one of the dominant controls on Snowball initiation and deglaciation \citep{Pollard2005,Lewis2006,Abbot2010,Abbot2011,Yang2012}; the relatively large range of albedo, 0.4 to 0.7, strengthens this importance. In the present model, increasing \(\alpha_i\) shifts both transitions toward substantially larger greenhouse forcing than the other parameter variations considered here.


Ice albedo affects not only the ice-line hysteresis curve (Fig.~\ref{fig:albedo_intercomparison}a), but also the other variables in Fig.~\ref{fig:albedo_intercomparison}. Higher albedo promotes larger ice volumes and therefore smaller ocean volumes, leading to higher salt concentration during snowball states. In contrast, the partial-ice-cover branches have nearly identical mean salinity and ocean depth across the albedo cases. Marked changes in these quantities occur only after the transition to a full-snowball, when a much larger fraction of the water column is stored in ice. The ice-line hysteresis confirms that high ice albedo strongly resists deglaciation and favors more extensive glaciation under the same external forcing.

Across the tested albedo range, initial salinity (Fig.~\ref{fig:salinity_comparison_subplots}) produces only a modest shift in the deglaciation threshold compared with the much larger shift caused by ice albedo (Fig.~\ref{fig:albedo_intercomparison}). Higher initial salinity lowers the seawater freezing temperature and therefore reduces the $F_{\mathrm{CO_2}}$ required for deglaciation. The $0.1~\mathrm{W\,m^{-2}}$ forcing increments used near the transitions resolve these small differences in Figs.~S2 and S3. A quantitative summary of the albedo-sensitivity experiments is given in Table~\ref{tab:S_albedo_sensitivity}.

\begin{table}[htbp]
\centering
\caption{Ice-albedo sensitivity for \(S_i=35\,\mathrm{PPT}\), \(Q=1285\,\mathrm{W\,m^{-2}}\), and \(G=0.06\,\mathrm{W\,m^{-2}}\). Here, \(S_f\) denotes the global-mean final ocean salinity of the full-snowball equilibrium just before the deglaciation transition. Approximate CO$_2$ concentrations use the same inverted \citet{ByrneGoldblatt2014} relation as Table~\ref{tab:S_salinity_sensitivity}; deglaciation values above \(0.01\,\mathrm{bar}\) are extrapolated estimates.}
\label{tab:S_albedo_sensitivity}
\footnotesize
\setlength{\tabcolsep}{4pt}
\renewcommand{\arraystretch}{1.12}
\begin{tabular}{cccccccc}
\toprule
\(\alpha_i\) & \(\overline{h_i}\) & \(S_f\) & \(F_{\mathrm{init}}\)
& \(p\mathrm{CO}_{2,\mathrm{init}}\) & \(F_{\mathrm{deg}}\)
& \(p\mathrm{CO}_{2,\mathrm{deg}}\) & \(S_i/S_f\) \\
-- & \(\mathrm{m}\) & \(\mathrm{PPT}\) & \(\mathrm{W\,m^{-2}}\)
& \(\mathrm{bar}\) & \(\mathrm{W\,m^{-2}}\) & \(\mathrm{bar}\) & -- \\
\midrule
0.4 & 1029.8 & 44.32 & -4.4 & \(1.16\times10^{-4}\) & 31.5 & 0.0243 & 0.79 \\
0.5 & 1242.9 & 47.19 & -1.7 & \(2.02\times10^{-4}\) & 36.6 & 0.0427 & 0.74 \\
0.6 & 1457.8 & 50.48 & 0.5 & \(3.07\times10^{-4}\) & 43.7 & 0.0899 & 0.69 \\
0.7 & 1734.6 & 55.46 & 2.5 & \(4.41\times10^{-4}\) & 50.8 & 0.182 & 0.63 \\
\bottomrule
\end{tabular}
\end{table}

Although salinity is the main focus of the present paper, the albedo comparison provides the broader parameter context in which the salinity effect should be interpreted. The sensitivity experiments confirm that surface albedo remains the most effective control on the glaciation and deglaciation thresholds in this class of energy-balance sea-glacier models, consistent with the central role of the ice--albedo feedback in Snowball Earth dynamics. This is physically expected, because changing the ice albedo directly modifies the absorbed shortwave radiation over the ice-covered domain and therefore strongly affects the stability of both the partially ice-covered and globally glaciated states.


\subsection{Relative importance of the varied model parameters}

To place the salinity effect in the context of other uncertain Neoproterozoic boundary conditions, we ran a full-factorial ensemble spanning three initial salinities (15, 25, and 35~PPT), four ice-surface albedos (0.4, 0.5, 0.6, and 0.7), three solar constants (1285, 1292, and 1299~$\mathrm{W\,m^{-2}}$), and two geothermal heat flux values (0.06 and 0.1~$\mathrm{W\,m^{-2}}$), yielding $3\times4\times3\times2=72$ cases in total (Supplementary Table~S1). For each case, we calculated the steady states and hysteresis curve of the system, recording the CO$_2$ forcing at Snowball initiation ($F_{\mathrm{init}}$) and at deglaciation ($F_{\mathrm{deg}}$). The purpose of this ensemble is not to rank the parameters in general. The ranges were chosen to represent plausible Neoproterozoic conditions: ice albedos of 0.4--0.7 represent different ice surfaces, from darker bare or dusty ice to bright snow-covered ice; initial salinities of 15, 25, and 35~PPT bracket fresher-than-modern to modern oceans within the calibrated range of the linear freezing-point relation (Eq.~\eqref{eq:Tf_linear_main}), although full-snowball salinities reach about 56~PPT and therefore extrapolate that relation; the solar constants are 5.0--6.0\% below the present-day value of 1367~\(\mathrm{W\,m^{-2}}\), representing the fainter Neoproterozoic Sun \citep{Gough1981}; and the two geothermal heat fluxes span plausible basal heat supply beneath thick marine ice \citep{JaupartMareschal2007}.

To compare the relative importance of the four parameters, we used a marginal-effects approach. Let $j$ denote one of the four parameters and let $\mathcal{L}_j$ denote its set of sampled levels. For each level $\ell\in\mathcal{L}_j$, we calculated the marginal mean
\begin{equation}
\overline{Y}_{j,\ell}=\frac{1}{N_{j,\ell}}\sum_{m:x_{m,j}=\ell}Y_m,
\label{eq:marginal_mean}
\end{equation}
where $Y_m$ is either $F_{\mathrm{init}}$ or $F_{\mathrm{deg}}$ for ensemble member $m$, $x_{m,j}$ is the value of parameter $j$ in that member, and $N_{j,\ell}$ is the number of members for which $x_{m,j}=\ell$. Thus, the mean at a given level of parameter $j$ averages over all combinations of the other three parameters. Each albedo mean contains 18 cases, each initial-salinity and solar-constant mean contains 24 cases, and each geothermal-flux mean contains 36 cases. The \textit{marginal-effect span}, $E_j$, is the difference between the largest and smallest marginal means across the sampled levels of parameter $j$:
\begin{equation}E_j(Y)=\max_{\ell\in\mathcal{L}_j}\left(\overline{Y}_{j,\ell}\right)-\min_{\ell\in\mathcal{L}_j}\left(\overline{Y}_{j,\ell}\right),
\label{eq:marginal_span}
\end{equation}A larger span indicates a stronger influence on the threshold. We then normalized these spans into percentage weights, $w_j$, so that the four parameter weights sum to 100\%:
\begin{equation}
w_j(Y)=100\,\frac{E_j(Y)}{\displaystyle\sum_{k=1}^{4}E_k(Y)}.
\label{eq:sensitivity_weight}
\end{equation}The sign of $\overline{Y}_{j,\ell_{\mathrm{high}}}-\overline{Y}_{j,\ell_{\mathrm{low}}}$, where $\ell_{\mathrm{high}}$ and $\ell_{\mathrm{low}}$ are the numerically highest and lowest sampled levels of parameter $j$, indicates whether increasing that parameter raises or lowers the threshold. The overall weight reported in Fig.~\ref{fig:sensitivity_weightage} is the arithmetic mean of the initiation and deglaciation weights for each parameter, based on the values in Supplementary Table~S1.

These weights should be interpreted as normalized threshold shifts over the sampled parameter ranges: a larger value indicates that varying that parameter produces a larger change in $F_{\mathrm{init}}$ or $F_{\mathrm{deg}}$ within the ensemble. They are not statistical probabilities or uncertainty estimates, and the analysis uses the directly imposed $F_{\mathrm{CO_2}}$ thresholds rather than the derived and partly extrapolated $p\mathrm{CO_2}$ concentrations. Because the weights scale with the sampled ranges, they describe the size of the salinity effect relative to the other controls over these Neoproterozoic-relevant ranges rather than a general ranking.

Ice albedo is the dominant control on both transition thresholds, with relative weights of $69.1\%$ for $F_{\mathrm{init}}$ and $80.5\%$ for $F_{\mathrm{deg}}$ (Fig.~\ref{fig:sensitivity_weightage}). Initial salinity produces the next-largest shifts for both thresholds, $21.3\%$ and $9.7\%$, respectively, followed by the solar constant and geothermal heat flux. The $0.1~\mathrm{W\,m^{-2}}$ forcing increments used near the transitions are fine enough to resolve these underlying threshold differences. Increasing ice albedo raises both the initiation and deglaciation thresholds, whereas increasing initial salinity or the solar constant lowers both. Geothermal heat flux has almost no marginal effect on initiation over the tested range but lowers the deglaciation threshold.

The supplementary material extends this comparison by showing additional model variables, including atmospheric humidity, precipitation and evaporation rates, freezing temperature, and the solar-constant and geothermal-heat-flux sensitivity tests for the full 72-case ensemble (Figs.~S1--S13). These diagnostics provide the broader context for the main result: salinity is not a passive tracer, but a coupled state variable that influences Snowball thresholds through freezing-point depression and ocean-volume change.

\section{Discussion}
\label{sec:discussion}



\begin{figure}[htbp]
    \centering
    \includegraphics[width=0.95\textwidth]{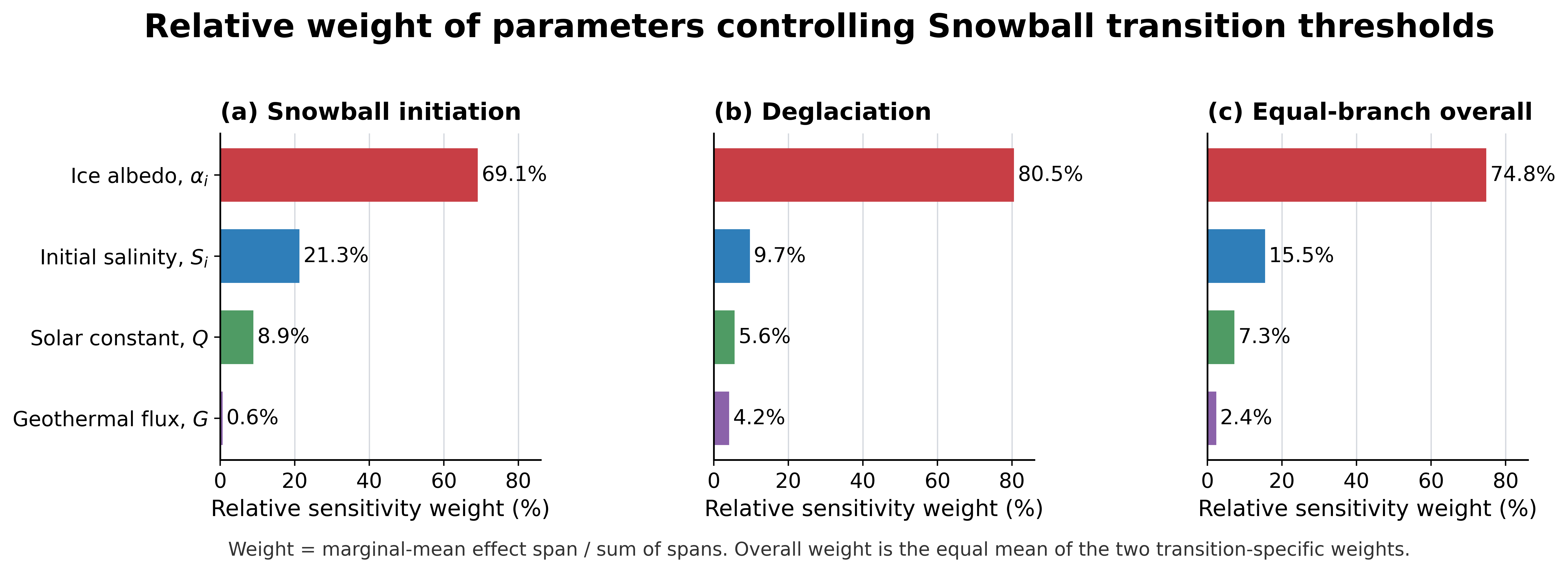}
    \caption{Relative marginal-effect weights of ice albedo, initial salinity, solar constant, and geothermal heat flux for (a) Snowball initiation, (b) deglaciation, and (c) their equal-branch mean. For each transition, the weight of a parameter is its marginal-mean threshold span across the sampled levels divided by the sum of the four parameter spans. The analysis uses the imposed $F_{\mathrm{CO_2}}$ thresholds rather than the derived $p\mathrm{CO_2}$ concentrations. The weights are specific to the parameter ranges sampled in this study.}
    \label{fig:sensitivity_weightage}
\end{figure}

The salinity experiments show a coupled thermodynamic and volume effect: higher salinity lowers the freezing temperature, while marine-ice growth reduces ocean volume and concentrates the conserved salt inventory. This process modifies both the hysteresis branches and their transition points. Over the sampled Neoproterozoic ranges, the marginal-effect analysis (Fig.~\ref{fig:sensitivity_weightage}) shows that the salinity effect is smaller than that of ice albedo but larger than those of the solar constant and geothermal heat flux for both thresholds. The paired salinity and ocean-depth curves make this mechanism explicit: as the system moves from open ocean toward full glaciation, the ocean reservoir shrinks, the remaining ocean becomes saltier, the freezing temperature decreases, and the full-snowball branch becomes thermally and compositionally distinct from the open-ocean branch.

The constant-\(\rho_o\) used in this study should therefore be read as a reduced-model approximation, not as a statement that salinity does not affect seawater density. High-salinity density measurements and seawater equation-of-state fits show that density increases measurably with salinity \citep{MilleroPoisson1981,MilleroHuang2009}. Relative to an initial salinity of 35~PPT, the increase to the highest full-snowball salinity reached here (about 56~PPT) raises the near-freezing density by approximately \(17~\mathrm{kg\,m^{-3}}\), or about \(1.7\%\) of seawater density. This percent-level correction could affect density-driven circulation or steric volume changes in a higher-dimensional ocean model, but those processes are outside the present slab-ocean model. In the present framework, salinity acts through freezing-point depression and the global conservation of salt and ocean mass; incorporating full equation-of-state and buoyancy feedbacks is left for future work.

The approximate conversion from $F_{\mathrm{CO_2}}$ to $p\mathrm{CO_2}$ concentration should be interpreted cautiously. The conversion used here is based on the high-CO$_2$ radiative-forcing formulation of \citet{ByrneGoldblatt2014}, but this is not intended to provide an exact reconstruction of atmospheric CO$_2$ at the very high forcing values relevant to snowball deglaciation, especially when extrapolated beyond its calibrated range. In the present study, it is therefore used only as a qualitative guide for interpreting the physical meaning of the forcing differences. This distinction is important because a difference of only about $1~\mathrm{W\,m^{-2}}$ in radiative forcing may appear modest, but in the high-$p\mathrm{CO_2}$ concentration regime it can correspond to a large difference in the CO$_2$ concentration required for deglaciation. Thus, when the results are viewed in terms of approximate $p\mathrm{CO_2}$ concentrations rather than radiative forcing alone, the salinity dependence of the deglaciation threshold becomes more physically significant: saltier oceans require distinctly lower CO$_2$ concentrations to escape the snowball state than fresher oceans.

The albedo experiments should be interpreted as sampling uncertainty in the effective optical state of Snowball ice rather than four discrete physical ice types. Snow cover, impurities, melt, brine, bubbles, grain-size evolution, and deformation can all alter albedo \citep{Warren2002,WarrenBrandt2008}. A spatially and temporally varying albedo treatment is beyond the present model. Within the explored range, however, the large hysteresis response demonstrates that albedo is the dominant control on the transition thresholds.

Across the full 72-case ensemble, the rapid final collapse begins when the ice line reaches $16.7^\circ$--$17.3^\circ$ latitude (approximately $17^\circ$). The remaining open-ocean belt then closes over the following $0.5~\mathrm{W\,m^{-2}}$ decrease in $F_{\mathrm{CO_2}}$. This narrow latitude range is evident in the ice-line trajectories in Fig.~\ref{fig:albedo_intercomparison}a and Figs.~S2--S3. Its existence is consistent with the classical interpretation of Snowball Earth initiation in energy-balance and climate models, where ice margins that advance sufficiently far into the tropics become unstable to runaway glaciation unless a stabilizing mechanism, such as a waterbelt or Jormungand-like state, arrests the feedback \citep{Budyko1969,Sellers1969,Hyde2000,Abbot2011,Yang2012,Pierrehumbert2011}. The numerical value of this limit depends on the model. Still, the underlying physical mechanism is the same: once ice expands into low latitudes, the increase in planetary albedo becomes strong enough that the remaining open ocean can no longer maintain a stable partial-ice state.

The spherical stress-divergence correction of \citet{Tziperman2012} (the \(-\cot^2\theta\,Bv\) term in Eq.~\eqref{eq:momentum_epsl}) arises because meridional convergence changes the membrane-stress divergence even in a latitude-only model; the numerical domain therefore excludes the singular pole and uses \(v=0\) at the boundary. Over the parameter range explored here, this correction changes the geometric redistribution of floating ice but does not substantially shift the ice-thickness scale or the hysteresis thresholds.

The supplementary figures (Figs.~S2--S13) for solar constant and geothermal heat flux place the salinity results in a broader physical context. Variations in solar constant from 1285 to \(1299~\mathrm{W\,m^{-2}}\) represent plausible Neoproterozoic insolation differences spanning the commonly adopted reduced solar luminosity, while geothermal heat-flux variations represent uncertainty in the basal internal heat supply of the early Earth. Figure~\ref{fig:sensitivity_weightage} places the salinity effect in context: over the sampled ranges, only ice albedo shifts the thresholds more than initial salinity, with equal-branch weights of \(74.8\%\), \(15.5\%\), \(7.3\%\), and \(2.4\%\) for ice albedo, initial salinity, solar constant, and geothermal heat flux, respectively. These weights scale with the sampled ranges and are not intended as a general ranking. 

Three simplifications limit the quantitative use of these thresholds. First, the model has no carbon cycle: \(F_{\mathrm{CO_2}}\) is imposed, so the results identify the forcing needed to cross each threshold, but not whether or how quickly volcanic outgassing and reduced weathering could supply it. Second, clouds are not represented explicitly, so their radiative feedbacks, which can alter Snowball thresholds substantially \citep{Pierrehumbert2011}, are absent. Third, the CO$_2$ forcing is spatially uniform, whereas the actual forcing depends on temperature and humidity and is weaker over cold, dry regions; a forcing that varies with latitude could shift the thresholds.

The elevated full-snowball salinities, \(40.0\)--\(55.9~\mathrm{PPT}\) across the experiments initialized from \(S_i=35~\mathrm{PPT}\) and \(50.5~\mathrm{PPT}\) in the reference case, are high but not by themselves prohibitive for life: hypersaline brines host active microbial communities today, and halotolerant and moderately halophilic microorganisms tolerate substantially higher salinities \citep{Murray2012,Cooper2019,Oren2015,Lee2018,Oren2024}. Light availability beneath thick, optically complex ice is likely a stronger limitation \citep{McKay2000,Warren2002}, restricting photosynthesis to refugia such as thin or clear ice, brine channels, polynyas, melt ponds, or restricted basins \citep{VincentHowardWilliams2000,HoffmanSchrag2002,Abbot2013,Campbell2014,Hawes2018,Song2023,Husain2025}, whereas non-photosynthetic ecosystems could persist through heterotrophy and chemolithoautotrophy supported by hydrothermal or water--rock inputs \citep{Christner2014,Lechte2019}. Salinity amplification therefore acts as an ecological filter rather than a sterilizing condition, shifting the snowball ocean toward cold, saline, low-light habitats.

\section{Summary}
\label{sec:summary}

Our goal was to determine how interactive ocean salinity affects Snowball Earth equilibria and transition thresholds, including states with partially ice-covered oceans. To address this question, we developed a one-dimensional, latitude-dependent coupled atmosphere--ocean--sea-glacier model with prognostic ocean salinity, global conservation of ocean mass and salt, and a salinity-dependent freezing temperature. The sea-glacier formulation retains the spherical stress-divergence correction of \citet{Tziperman2012} to the \citet{Pollard2005} momentum balance; this term mainly affects how floating ice thickness is redistributed, while the bifurcation thresholds remain controlled primarily by radiative forcing, albedo, and the thermodynamic ice--ocean balance. The model produced two stable climate equilibria under the same imposed CO$_2$ radiative forcing, $F_{\mathrm{CO_2}}$: a partially ice-covered ocean state and a globally ice-covered full-snowball state. These equilibria differed strongly in ice thickness, ocean volume, salinity, and thermal structure. In the full-snowball branch, the growth of thick floating ice reduced the ocean volume and concentrated salt, producing a colder and more saline ocean.

The sensitivity experiments show that ice-surface albedo remains the dominant control on snowball initiation and deglaciation, consistent with the central role of the ice--albedo feedback. However, salinity emerges as an important internal control on the coupled ice--ocean system. Increasing the initial salinity lowers the seawater freezing temperature, shifts both initiation and deglaciation toward lower CO$_2$ forcing, and substantially modifies the final snowball ocean salinity. Although the shift in radiative forcing is modest, its implication for the required high-$p\mathrm{CO_2}$ concentration for deglaciation can be significant. Across the explored parameter range, salinity affects not only the transition thresholds but also the physical character of the snowball ocean through its coupling to ice thickness and ocean volume.


The broader parameter survey further shows that the solar constant and geothermal heat flux affect the transition thresholds in the expected direction, but their relative weights are smaller than that of initial salinity over the ranges tested here (Fig.~\ref{fig:sensitivity_weightage}).
Overall, the results suggest that salinity should not be treated as a passive tracer in Snowball Earth models. Instead, it is a dynamically important state variable that links ice growth, ocean-volume reduction, freezing temperature, deglaciation behavior, and the potential habitability of cold, hypersaline snowball oceans.

The role of salinity in Snowball Earth dynamics carries broader implications for icy moons and tidally locked exoplanets, where similar ice–ocean salinity feedbacks may operate. By lowering the freezing point of the underlying ocean, salinity dynamics could shift the boundaries of the habitable zone, suggesting that future assessments of planetary and satellite habitability should explicitly account for salinity-dependent freezing thresholds rather than treating the ocean as pure water.

\appendix


\section{Model equations, parameters, and numerical implementation}
\setcounter{table}{0}


This appendix summarizes the governing equations, parameter definitions, forcing terms, and numerical implementation of the coupled one-dimensional atmosphere--ocean--sea-glacier model. The climate component follows the framework of \citet{Pollard2005}, while the sea-glacier dynamics use the one-dimensional spherical formulation derived from \citet{Tziperman2012}. The subgrid migration of the ice front follows the fractional-cover approach of \citet{Albrecht2011}. Unless otherwise stated, parameter values are those of \citet{Pollard2005}. The two deliberate modifications used in the present study are the ocean albedo, prescribed as \(\alpha_o=0.10\), and the ice--ocean heat-exchange coefficient, set to \(\beta=400~\mathrm{W\,m^{-2}\,K^{-1}}\) following \citet{Ashkenazy2018,HollandJenkins1999}.


\subsection{Variables, parameters, and forcing terms}


\subsubsection{Prognostic and diagnostic variables}
The main prognostic and diagnostic variables of the model are listed in Table~\ref{tab:variables}.

\begin{table}[H]
\centering
\caption{Main prognostic and diagnostic variables of the model.}
\small\setstretch{1}
\label{tab:variables}
\begin{tabular}{>{\raggedright\arraybackslash}p{0.24\textwidth}>{\raggedright\arraybackslash}p{0.66\textwidth}}
\hline
Symbol & Description \\
\hline
\(T_o(\theta,t)\) & Ocean temperature (K) \\
\(T_a(\theta,t)\) & Atmospheric temperature (K) \\
\(q_a(\theta,t)\) & Atmospheric specific humidity (dimensionless) \\
\(T_s(\theta,t)\) & Ice-surface temperature (K) \\
\(S_o(\theta,t)\) & Local ocean salinity (PPT) \\
\(\langle S_o \rangle(t)\) & Globally averaged ocean salinity (PPT) \\
\(h_i(\theta,t)\) & Floating ice thickness (m) \\
\(v(\theta,t)\) & Meridional ice velocity (\(\mathrm{m\,s^{-1}}\)) \\
\(h_o(t)\) & Globally averaged effective ocean thickness (m) \\
\(T_f(\theta,t)\) & Local freezing temperature at the ice--ocean interface (K) \\
\(R(\theta,t)\) & Fractional ice-cover field used in the subgrid front parameterization (dimensionless) \\
\hline
\end{tabular}
\end{table}

\subsubsection{Physical constants and transport parameters}
Table~\ref{tab:parameters} lists the physical constants and transport parameters used in the model.

{\small\setstretch{1}
\begin{longtable}{>{\raggedright\arraybackslash}p{0.15\textwidth}>{\raggedright\arraybackslash}p{0.45\textwidth}>{\raggedright\arraybackslash}p{0.3\textwidth}}
\caption{Physical constants and transport parameters.}
\label{tab:parameters}\\
\hline
Symbol & Description & Value used \\
\hline
\endfirsthead
\caption[]{Physical constants and transport parameters (continued).}\\
\hline
Symbol & Description & Value used \\
\hline
\endhead
\(r\) & Earth radius (m) & \(6.3\times10^{6}\) \\
\(g\) & Gravitational acceleration (\(\mathrm{m\,s^{-2}}\)) & \(9.8\) \\
\(\rho_a\) & Air density (\(\mathrm{kg\,m^{-3}}\)) & \(1.214\) \\
\(\rho_o\) & Seawater reference density (\(\mathrm{kg\,m^{-3}}\)) & \(1000\) \\
\(\rho_i\) & Ice density (\(\mathrm{kg\,m^{-3}}\)) & \(900\) \\
\(c_a\) & Specific heat of air (\(\mathrm{J\,kg^{-1}\,K^{-1}}\)) & \(1004.64\) \\
\(c_o\) & Specific heat of seawater (\(\mathrm{J\,kg^{-1}\,K^{-1}}\)) & \(4218\) \\
\(\sigma\) & Stefan--Boltzmann constant (\(\mathrm{W\,m^{-2}\,K^{-4}}\)) & \(5.66961\times10^{-8}\) \\
\(\kappa_i\) & Thermal conductivity of ice (\(\mathrm{W\,m^{-1}\,K^{-1}}\)) & \(2.1\) \\
\(L_f,L_v,L_s\) & Latent heats of fusion, vaporization, and sublimation (\(\mathrm{J\,kg^{-1}}\)) & \(3.34\times10^{5},\ 2.5104\times10^{6},\ 2.8440\times10^{6}\) \\
\(D_a\) & Meridional diffusivity of atmospheric heat (\(\mathrm{m^2\,s^{-1}}\)) & \(1.3\times10^{6}\) \\
\(D_o\) & Meridional diffusivity of ocean heat and salinity (\(\mathrm{m^2\,s^{-1}}\)) & \(2.5\times10^{2}\) \\
\(D_q\) & Meridional diffusivity of atmospheric moisture (\(\mathrm{m^2\,s^{-1}}\)) & \(1.69\times10^{6}\) \\
\(\kappa\) & Numerical diffusion coefficient in the ice-thickness equation (\(\mathrm{m^2\,s^{-1}}\); Eq.~\eqref{eq:ice-thickness1}) & \(1\) \\
\(\beta\) & Ice--ocean heat-exchange coefficient (\(\mathrm{W\,m^{-2}\,K^{-1}}\)) & \(400\) \\
\(\tau_p\) & Precipitation timescale (s) & \(1.0368\times10^{6}\) (12 days) \\
\(C_D f_\theta\) & Bulk transfer velocity for sensible and latent fluxes (\(\mathrm{m\,s^{-1}}\)) & \(1.07\times10^{-3}\) (sensible); \(7.93\times10^{-3}\) (latent) \\
\(\epsilon\) & Atmospheric longwave emissivity (dimensionless) & \(0.70\) \\
\(R_d/R_v\) & Ratio of the gas constants of dry air and water vapor, used for saturation specific humidity (dimensionless) & \(0.622\) \\
\(h_a\) & Effective atmospheric thickness for heat (m) & \(8400\) \\
\(h_q\) & Effective atmospheric thickness for water vapor (m) & \(6000\) \\
\(\alpha_a\) & Atmospheric shortwave albedo (dimensionless) & \(0.24\) \\
\(\tau_a\) & Atmospheric shortwave transmissivity (dimensionless) & \(0.63\) \\
\(c_i\) & Specific heat of ice (\(\mathrm{J\,kg^{-1}\,K^{-1}}\)) & \(2106\) \\
\(n\) & Glen's flow-law exponent (dimensionless) & \(3\) \\
\(\alpha_o\) & Ocean albedo (dimensionless) & \(0.1\) \\
\hline
\end{longtable}}

\subsubsection{Control parameters and external forcing}
Table~\ref{tab:forcings} summarizes the control parameters varied in the sensitivity experiments and the imposed CO$_2$ forcing.

\begin{table}[H]
\centering
\caption{Main control parameters explored in the sensitivity experiments.}
\small\setstretch{1}
\label{tab:forcings}
\begin{tabular}{>{\raggedright\arraybackslash}p{0.16\textwidth}>{\raggedright\arraybackslash}p{0.46\textwidth}>{\raggedright\arraybackslash}p{0.28\textwidth}}
\hline
Symbol & Description & Values or range used \\
\hline
\(\alpha_i\) & Ice-surface albedo & 0.4, 0.5, 0.6, 0.7 \\
\(S_i\) & Initial mean ocean salinity & 15, 25, 35~PPT \\
\(Q\) & Solar constant & 1285, 1292, 1299~\(\mathrm{W\,m^{-2}}\) \\
\(G\) & Geothermal heat flux & 0.06, 0.10~\(\mathrm{W\,m^{-2}}\) \\
\(F_{\mathrm{CO_2}}\) & Imposed longwave \(\mathrm{CO_2}\) radiative forcing & varied along each hysteresis branch \\
\hline
\end{tabular}
\end{table}

\subsection{Atmosphere--ocean equations}
\label{app:thermodynamic_equations}

\subsubsection{Meridional diffusion operators}

All climate equations are solved on a one-dimensional spherical domain in co-latitude \(\theta\). The diffusion terms use the same indexed operator introduced in Eq.~\eqref{eq:diffusion_operator},

\begin{equation}
\mathcal{D}_{j}[X_j]
=
\frac{1}{r^{2}\sin\theta}
\frac{\partial}{\partial\theta}
\left(
D_j \sin\theta \frac{\partial X_j}{\partial\theta}
\right),
\label{eq:app_diffusion_operator}
\end{equation}
with \(j\in\{o,a,q,S\}\) and
\begin{equation}
(X_j,D_j)=
\begin{cases}
(T_o,D_o), & j=o,\\
(T_a,D_a), & j=a,\\
(q_a,D_q), & j=q,\\
(S_o,D_o), & j=S.
\end{cases}
\label{eq:app_diffusion_index_map}
\end{equation}
Thus the salinity operator uses the same ocean diffusivity \(D_o\) as the ocean-temperature operator.

\subsubsection{Atmospheric temperature}

The atmospheric temperature equation is
\begin{equation}
\rho_a c_a h_a \frac{\partial T_a}{\partial t}
=
\rho_a c_a h_a\,\mathcal{D}_{a}[T_a]
+
Q_a + I_a + H_a + LP + F_{\mathrm{CO_2}}.
\label{eq:app_Ta}
\end{equation}

where \(Q_a\), \(I_a\), and \(H_a\) are, respectively, atmospheric shortwave absorption, longwave radiative heating, and sensible heat exchange with the lower boundary; \(LP\) is latent heat released by precipitation. All four of these terms have units \(\mathrm{W\,m^{-2}}\). The imposed \(F_{\mathrm{CO_2}}\) is a spatially uniform positive longwave heating term, also with units \(\mathrm{W\,m^{-2}}\), added directly to the atmospheric energy budget. The compact forcing \(\mathcal{F}_a\) in Eq.~\eqref{eq:Ta_compact_main} is the sum of the four internally calculated terms; \(F_{\mathrm{CO_2}}\) is added separately as shown above.

\subsubsection{Atmospheric humidity}

The atmospheric specific-humidity equation is
\begin{equation}
\rho_a h_q \frac{\partial q_a}{\partial t}
=
\rho_a h_q\,\mathcal{D}_{q}[q_a]
-
(P-E),
\label{eq:app_qa}
\end{equation}
where \(P\) is precipitation and \(E\) is evaporation or sublimation, both with units \(\mathrm{kg\,m^{-2}\,s^{-1}}\). Thus \(P-E\) is the net atmospheric moisture loss; precipitation removes water vapor from the atmospheric column, while evaporation or sublimation returns water vapor to it.

\subsubsection{Ocean temperature}

The ocean-temperature equation is
\begin{equation}
\rho_o c_o h_o \frac{\partial T_o}{\partial t}
=
\rho_o c_o h_o\,\mathcal{D}_{o}[T_o]
+
Q_o + I_o + H_o - L E_o + G - \beta (T_o-T_f).
\label{eq:app_To}
\end{equation}

where \(Q_o\), \(I_o\), and \(H_o\) are the ocean shortwave, longwave, and sensible-heat terms, \(LE_o\) is latent heat loss by open-ocean evaporation, \(G\) is geothermal heat input, and \(\beta(T_o-T_f)\) is the ice--ocean heat-exchange term. These terms have units \(\mathrm{W\,m^{-2}}\). The subscript \(o\) indicates that \(E_o\) applies to exposed ocean; over ice-covered cells, the corresponding surface sublimation term is \(E_s\) in Eq.~\eqref{eq:app_Ts}.

\subsubsection{Ocean salinity}

The salinity equation is
\begin{equation}
h_o \frac{\partial S_o}{\partial t}
=
h_o\,\mathcal{D}_{S}[S_o]
+
\mathcal{B}_S,
\label{eq:app_So}
\end{equation}
where $\mathcal{B}_S$ is the local salt source/sink given by
\begin{equation}
\mathcal{B}_S
=
\langle S_o \rangle F_o
+
\langle S_o \rangle \frac{E}{\rho_o}
-
\langle S_o \rangle \frac{P}{\rho_o}.
\label{eq:app_BS}
\end{equation}
Here \(F_o\) is the rate of basal freezing or melting at the ice--ocean interface. It has units of \(\mathrm{m\,s^{-1}}\): positive \(F_o\) increases ice thickness and rejects salt to the ocean, whereas negative \(F_o\) represents basal melting and freshwater input. It is calculated from the local ice--ocean energy balance,
\begin{equation}
F_o
=
\frac{1}{L\rho_o}
\left[
\kappa_i\frac{T_f-T_s}{h_i}
-
\beta(T_o-T_f)
\right],
\label{eq:app_Fo}
\end{equation}
where the first term in brackets is conductive heat loss through the ice and the second is turbulent heat exchange with the ocean. Both bracketed terms have units \(\mathrm{W\,m^{-2}}\), so division by \(L\rho_o\) gives \(F_o\) units of \(\mathrm{m\,s^{-1}}\). The Pollard--Kasting precipitation and evaporation rates, \(P\) and \(E\), are divided by \(\rho_o\) to convert them from mass fluxes to equivalent ocean-depth tendencies.

\subsubsection{Ice-surface temperature}

Following \citet{Pollard2005}, the ice/snow surface-temperature equation is
\begin{equation}
C_s \frac{\partial T_s}{\partial t}
=
Q_s + I_s + H_s - L E_s + F_c,
\label{eq:app_Ts}
\end{equation}
where \(Q_s\), \(I_s\), and \(H_s\) are the shortwave, longwave, and sensible-heat terms at the ice/snow surface, \(LE_s\) is latent heat loss by sublimation from the ice/snow surface, and \(F_c\) is conductive heat supplied from below. These forcing terms have units \(\mathrm{W\,m^{-2}}\). The conductive flux through the ice column is
\begin{equation}
F_c = \kappa_i \frac{T_f-T_s}{h_i}.
\label{eq:app_Fc}
\end{equation}

\subsubsection{Freezing temperature}

The freezing point of seawater decreases with salinity and pressure; we use the standard linear liquidus approximation based on the seawater thermodynamic relation of \citet{Millero1978} and widely used in ice-shelf/ocean boundary-layer models \citep{HollandJenkins1999,HellmerOlbers1989,JenkinsHellmerHolland2001,Losch2008}. Expressed in Kelvin,
\begin{equation}
T_f =
273.15 + 0.0901 - 0.0575\,S_o - 7.61\times10^{-4}p_b,
\label{eq:app_Tf}
\end{equation}
where \(S_o\) is salinity in PPT and \(p_b\) is basal ice pressure in dbar. This freezing temperature couples the ocean-temperature equation, the salinity equation, and the ice thermodynamics.


\subsection{Flux formulation}
\label{app:flux_parameterizations}


The longwave atmospheric heating term follows the grey-atmosphere Stefan--Boltzmann balance,
\begin{equation}
I_a = \sigma T_s^4 - 2\epsilon \sigma T_a^4,
\label{eq:app_Ia}
\end{equation}
where \(T_s\) is the ice/snow surface temperature and \(T_a\) is the atmospheric temperature. The factor of 2 appears because the atmospheric layer emits thermal radiation both upward to space and downward toward the surface, while \(\epsilon\) is the atmospheric emissivity.

Following \citet{Pollard2005}, precipitation is represented as an e-folding removal of atmospheric column water vapor, with the relative-humidity reduction written here as
\begin{equation}
P = \frac{\rho_a h_q q_a}{\tau_p} r_a^3,
\label{eq:app_P}
\end{equation}
where \(r_a\) is atmospheric relative humidity. Since \(q_a\) and \(r_a\) are dimensionless, \(P\) has units \(\mathrm{kg\,m^{-2}\,s^{-1}}\).

Evaporation or sublimation is parameterized with the stability-corrected bulk-transfer formulation used by \citet{Pollard2005}, where the stability factor follows \citet{Louis1979} and the land-surface-transfer implementation of \citet{PollardThompson1995}:
\begin{equation}
E = \rho_a C_D f_\theta \Delta q,
\label{eq:app_E}
\end{equation}
where \(\Delta q\) is the humidity deficit relative to saturation over water or ice, evaluated over the exposed ocean for \(E_o\) and over the ice/snow surface for \(E_s\). With \(C_D\) and \(\Delta q\) dimensionless and \(f_\theta\) a velocity scale, \(E\) has units \(\mathrm{kg\,m^{-2}\,s^{-1}}\). These Pollard--Kasting mass fluxes enter latent-heat terms as \(LP\) or \(LE\), and enter the salinity budget as \(P/\rho_o\) and \(E/\rho_o\).

The sensible heat flux is represented as
\begin{equation}
H = \rho_a c_a C_D f_\theta \Delta T,
\label{eq:app_H}
\end{equation}
Here, \(\Delta T\) is the local near-surface air--surface temperature difference. For \(H_o\), the surface is the ocean; for \(H_s\), it is the snow--ice surface.

The shortwave terms \(Q_a\), \(Q_o\), and \(Q_s\) follow the partitioning used in \citet{Pollard2005}.


\subsection{Mass and salt closure}
\label{app:mass_salt_closure}


Because \(h_o\) is treated here as the mean ocean depth rather than the depth of a shallow mixed layer, the total water inventory of ocean plus ice is conserved when the water content of the atmosphere is neglected:
\begin{equation}
\rho_i \langle h_i \rangle + \rho_o h_o = C_w,
\label{eq:app_mass}
\end{equation}
where $C_w$ is the mean total water mass (ocean plus ice) per unit area, with units of \(\mathrm{kg\,m^{-2}}\). Here \(\langle h_i \rangle\) is the area-weighted global-mean ice thickness,
\begin{equation}
\langle h_i \rangle = \frac{1}{2}\int_{0}^{\pi} h_i(\theta,t)\sin\theta\,d\theta,
\label{eq:app_mean_hi}
\end{equation}
whereas \(h_i(\theta,t)\) denotes the local latitude-dependent ice thickness. Hence, the change in ocean mean depth as a result of changes in ice thickness is
\begin{equation}
\delta h_o = -\frac{\rho_i}{\rho_o}\,\delta\langle h_i \rangle.
\label{eq:app_dho}
\end{equation}

The total salt content of the ocean is also conserved through the asynchronous coupling step:
\begin{equation}
\langle S_o \rangle_i h_{o,i} = \langle S_o \rangle_f h_{o,f}.
\label{eq:app_salt}
\end{equation}
Here \(\langle S_o \rangle\) is the globally averaged ocean salinity and \(h_o\) is the globally averaged effective ocean thickness. The closure applies to these global-mean quantities, not pointwise to the latitude-dependent salinity field.


\subsection{Modified one-dimensional sea-glacier equation}
\label{app:sea_glacier_equation}


The sea-glacier stress-balance derivation is given by \citet{Tziperman2012}. Here we provide only the final one-dimensional spherical form used in the present model, together with the associated stress-coefficient notation and front boundary condition.

\subsubsection{Momentum equation}

The effective ice viscosity in the vertically integrated momentum balance is given by
\begin{equation}
B = \frac{h_i}{r}
    \left\langle A(T)^{-1/3} \right\rangle
    \dot{\epsilon}^{-2/3},
\label{eq:app_B}
\end{equation}
where $\dot{\epsilon}$ is the effective one-dimensional spherical strain-rate invariant, $A(T)$ is the temperature-dependent Arrhenius rate factor, and $\langle \cdot \rangle$ denotes vertical averaging. Thus, $B$ is the depth-integrated effective viscosity entering the
membrane-stress balance.

The final one-dimensional spherical momentum equation is
\begin{equation}
\begin{split}
&\frac{1}{\sin\theta}
\frac{\partial}{\partial\theta}
\left(
B \sin\theta \frac{\partial v}{\partial\theta}
\right)
+
\frac{\partial}{\partial\theta}
\left[
\frac{B}{\sin\theta}
\frac{\partial}{\partial\theta}
\left(
v\sin\theta
\right)
\right]
\\
&\quad -
\cot^2\theta\,Bv
-
g\rho_i(1-\mu)h_i\frac{\partial h_i}{\partial\theta}
= 0,
\end{split}
\label{eq:app_mom}
\end{equation}
where \(\theta\) is colatitude, \(v\) is the ice meridional velocity, \(g\) is the gravitational acceleration, \(\rho_i\) is the ice density, \(\mu=\rho_i/\rho_o\) is the ice-to-seawater reference-density ratio, and \(h_i\) is the ice thickness. The first two terms are the spherical membrane-stress divergence, the \(-\cot^2\theta\,Bv\) term is the curvature correction, and the final term is the hydrostatic driving stress from gradients in ice thickness.

\subsubsection{Ice thickness equation}
Following \citet{Tziperman2012}, the corresponding local ice-thickness evolution equation is
\begin{equation}
    \frac{\partial h_i}{\partial t}+\frac{1}{r \sin{\theta}}\frac{\partial}{\partial \theta}(v h_i\sin{\theta})=\kappa\nabla^2h_i+S(\theta,t),
    \label{eq:ice-thickness1}
\end{equation}
where the left-hand side represents local thickening plus the divergence of meridional ice flux. The diffusion term is included only for numerical regularization; the coefficient \(\kappa\) is therefore chosen as small as stable integration permits. The source term \(S(\theta,t)\), with units of \(\mathrm{m\,s^{-1}}\), is the net local ice-thickness tendency from thermodynamics: basal freezing or melting at the ice--ocean interface is represented by \(F_o\) in Eq.~\eqref{eq:app_Fo}, while surface accumulation or ablation is associated with precipitation and evaporation or sublimation.

\subsubsection{Terminus force balance}

At the ice front, the depth-integrated force balance is
\begin{equation}
B\left[v\cot(\theta)+2\frac{\partial v}{\partial\theta}\right]
=
\frac{1}{2}g\rho_i h_i^2 (1-\mu).
\label{eq:app_front}
\end{equation}


\subsection{Subgrid ice-front parameterization}
\label{app:subgrid_front}


Because the sea-glacier equation is solved here on a fixed one-dimensional latitude grid following the finite-difference formulation of \citet{Tziperman2012}, the ice front does not generally coincide with a grid point. The fractional variable \(R\) is therefore used as a numerical bookkeeping device for the moving terminus, rather than as an additional physical closure. The model represents the front cell by the fractional ice-cover field
\begin{equation}
R =
\begin{cases}
0, & h_j = 0,\\[4pt]
h_j/H_{\mathrm{cr}}, & 0 < h_j < H_{\mathrm{cr}},\\[4pt]
1, & h_j \ge H_{\mathrm{cr}},
\end{cases}
\label{eq:app_R}
\end{equation}
where $h_j$ is the ice thickness at grid point $j$. Open ocean corresponds to $R=0$, a partially ice-covered front cell has $0<R<1$, and a fully ice-covered grid cell has $R=1$. The threshold thickness is taken to be $H_{\mathrm{cr}}=1~\mathrm{m}$.

The subgrid update applies only to the grid cell adjacent to the active ice front. When \(0<R<1\), the advective ice flux from the neighboring ice-covered grid point is distributed over the fractional cell width \(R\Delta\theta_{\mathrm{rad}}\), where \(\Delta\theta_{\mathrm{rad}}\) is the meridional grid spacing in radians. In the Northern Hemisphere, ice flowing from grid point \(j\) into the neighboring front cell changes the thickness at \(j+1\) according to
\begin{equation}
\left.\frac{\partial h_i}{\partial t}\right|_{j+1}
=
S(\theta_{j+1},t)
+
\frac{v_j h_j}{R \Delta\theta_{\mathrm{rad}}},
\label{eq:app_SC_NH}
\end{equation}
where \(S(\theta,t)\) is the local source term (Eq.~\eqref{eq:ice-thickness1}). In the Southern Hemisphere, the corresponding front cell lies at \(j-1\), giving
\begin{equation}
\left.\frac{\partial h_i}{\partial t}\right|_{j-1}
=
S(\theta_{j-1},t)
-
\frac{v_j h_j}{R \Delta\theta_{\mathrm{rad}}}.
\label{eq:app_SC_SH}
\end{equation}
The opposite signs of the advective terms reflect the opposite meridional orientation of the active front in the two hemispheres.


\subsection{Numerical implementation}
\label{app:numerical_implementation}


\subsubsection{Adams--Bashforth time stepping}

All prognostic climate and thermodynamic equations are integrated using second-order Adams--Bashforth:
\begin{equation}
X_j^{n+2}
=
X_j^{n+1}
+
\frac{\Delta t}{2}
\left[
3\,\mathrm{RHS}_j^{\,n+1}
-
\mathrm{RHS}_j^{\,n}
\right],
\label{eq:app_AB2}
\end{equation}
where \(X\) represents \(T_o\), \(T_a\), \(q_a\), \(T_s\), \(S_o\), or \(h_i\).

\subsubsection{Ice-thickness discretization at the terminus}
The terminus requires special numerical treatment; the corresponding discretization is described below.
Using Eq. \eqref{eq:ice-thickness}, at the Northern Hemisphere terminus, the advection term is discretized using backward differencing:
\begin{equation}
\left.
\frac{1}{r\sin\theta}
\frac{\partial}{\partial\theta}
(\sin\theta\,v h_i)
\right|_j
=
\frac{1}{r_j\sin\theta_j}
\frac{\sin\theta_j v_j h_j - \sin\theta_{j-1} v_{j-1} h_{j-1}}{\Delta\theta}.
\label{eq:app_adv_nh}
\end{equation}
At the Southern Hemisphere terminus, the advection term is discretized using forward differencing:
\begin{equation}
\left.
\frac{1}{r\sin\theta}
\frac{\partial}{\partial\theta}
(\sin\theta\,v h_i)
\right|_j
=
\frac{1}{r_j\sin\theta_j}
\frac{\sin\theta_{j+1} v_{j+1} h_{j+1} - \sin\theta_j v_j h_j}{\Delta\theta}.
\label{eq:app_adv_sh}
\end{equation}

\subsubsection{Tridiagonalization of the momentum equation}

The momentum equation, Eq. \eqref{eq:app_mom}, is discretized into the tri-diagonal form
\begin{equation}
a_j v_{j-1} + b_j v_j + c_j v_{j+1} = d_j,
\label{eq:app_tri}
\end{equation}
where the coefficients are assembled from the finite-difference discretization of Eq.~\eqref{eq:app_mom}, following the one-dimensional spherical formulation in the original numerical derivation.

At each coupling step:
\begin{enumerate}
    \item the effective ice-viscosity field \(B_j\) is initialized from the previous iteration,
    \item the coefficients \(a_j\), \(b_j\), \(c_j\), and \(d_j\) are assembled,
    \item the resulting tridiagonal system is solved for \(v_j\),
    \item the strain-rate field and the effective ice viscosity \(B_j\) are updated,
    \item the procedure is iterated until convergence.
\end{enumerate}

\subsubsection{Asynchronous coupling}
The atmosphere--ocean component is integrated with the smaller time step using the Adams--Bashforth scheme above. Because the ice responds much more slowly, the ice-thickness equation is advanced with a longer time step using the diagnostic velocity solution. The updated ice thickness then determines a new ocean depth through Eq.~\eqref{eq:app_dho}, after which the global salinity is adjusted accordingly through Eq.~\eqref{eq:app_salt}. These updated fields are then passed back to the atmosphere--ocean equations, closing the loop for the next coupling interval.

\subsection{Model configuration and numerical settings}
\label{app:configuration}

The model is an aquaplanet without continents or rotation. The meridional domain extends from 80\(^\circ\)S to 80\(^\circ\)N (co-latitudes 10\(^\circ\)--170\(^\circ\)) and is discretized with 89 latitude points (87 interior points and two boundary points), giving a grid spacing of about \(1.86^\circ\). Insolation is prescribed as an annual-mean meridional distribution proportional to \(1+s_2P_2(\cos\theta)\), where \(P_2\) is the second Legendre polynomial and \(s_2=-0.482\) \citep{Abbot2011}, normalized by its mean over the model domain. The atmosphere--ocean equations are integrated with a time step of 15--30~min for 20--40 model years in each coupling cycle, and the sea-glacier equations with a time step of 1~yr for \(10^6\)~yr in each coupling cycle; each equilibrium is reached after 20--50 coupling cycles. The ocean is initialized with a depth of 4000~m, and simulations started from the fully ice-covered state use an initial ice thickness of 1500~m. The remaining parameter values are listed in Table~\ref{tab:parameters}.






\bibliographystyle{elsarticle-harv}
\section*{CRediT authorship contribution statement}
Sounav Sengupta: Conceptualization, Methodology, Software, Formal analysis, Investigation, Visualization, Writing -- original draft; Francisco E. Spaulding-Astudillo: Conceptualization, Methodology, Software; Dorian S. Abbot: Conceptualization, Supervision, Writing -- review \& editing; Eli Tziperman: Conceptualization, Supervision, Funding acquisition, Writing -- review \& editing; Yosef Ashkenazy: Conceptualization, Methodology, Formal analysis, Supervision, Writing -- review \& editing.

\section*{Acknowledgements}
The authors thank Mahidip Deb, Sayan Sen, and Jeremie Schmiedel for their valuable insights and Joydeep Sinha for computational assistance. ET thanks the Weizmann Institute for its hospitality during parts of this work.

\noindent\textbf{Funding:} This work was supported by the National Science Foundation, Office of Polar Programs [grant number 2518542] (E.T.). The funder had no role in study design; in the collection, analysis and interpretation of data; in the writing of the report; or in the decision to submit the article for publication.

\section*{Declaration of competing interest}
The authors declare that they have no known competing financial interests or personal relationships that could have appeared to influence the work reported in this paper.

\section*{Data and code availability}
No observational data were used in this study. The model code used for all simulations is publicly available at \url{https://github.com/sounav342/snowball-earth-climate-ice-model}. The transition thresholds for all 72 experiments are listed in Supplementary Table~S1.

\section*{Declaration of generative AI and AI-assisted technologies in the manuscript preparation process}
During the preparation of this work, the authors used OpenAI Codex to assist with language editing, organization, and reference checks. After using this tool, the authors reviewed and edited the content as needed and take full responsibility for the content of the published article.

\bibliography{references}

\end{document}